\documentclass[12pt,preprint]{aastex}

\slugcomment{Accepted for publication in AJ}

\shorttitle{Brown Dwarfs and Very Low Mass Stars in Orion OB1}
\shortauthors{Downes et al.}

\begin{document}

\title{A large-scale optical-near infrared survey for brown dwarfs and very low-mass stars in the Orion OB1 association}

\author{Juan Jos\'e Downes W.}
\affil{Centro de Investigaciones de Astronom{\'\i}a, Apartado Postal. 264, M\'erida
5101-A, Venezuela}
\affil{Escuela de F{\'\i}sica, Universidad Central de Venezuela, Apartado Postal
47586, Caracas 1041-A, Venezuela}
\email{jdownes@cida.ve}

\author{C\'esar Brice\~no}
\affil{Centro de Investigaciones de Astronom{\'\i}a, Apartado Postal. 264, M\'erida
5101-A, Venezuela}
\email{briceno@cida.ve}

\author{Jes\'us Hern\'andez}
\affil{Department of Astronomy, University of Michigan, 825 Dennison Building, 500 Church Street, Ann Arbor, MI 48109} 
\affil{Centro de Investigaciones de Astronom{\'\i}a, Apartado Postal. 264, M\'erida 5101-A, Venezuela}
\email{hernandj@umich.edu}

\author{Nuria Calvet \& Lee Hartmann}
\affil{Department of Astronomy, University of Michigan, 825 Dennison Building, 500 Church Street, Ann Arbor, MI 48109 }
\email{ncalvet@umich.edu, lhartm@umich.edu}

\author{Ernesto Ponsot Balaguer}
\affil{Facultad de Ciencias Econ\'omicas y Sociales, Universidad de los Andes,
M\'erida, 5101, Venezuela}
\email{ernesto@ula.ve}

\author{ }

\begin{abstract}

We report the initial results of a large-scale optical-near infrared survey 
to extend the known young population of the entire Orion 
star-forming region down to the substellar domain. 
Using deep optical I-band photometry and data 
from the 2MASS survey, we selected candidates across $\sim$ $14.8$ $deg^2$ 
in the $\sim 8$ Myr old Ori OB1a subassociation and 
over $\sim$  $6.7$ $deg^2$ in the Ori OB1b subassociation (age $\sim 3$ Myr), 
with completeness down to $0.05M_{\odot}$ and $0.072M_{\odot}$ respectively.
We obtained low resolution optical spectra for a subsample 
of 4 candidates in Ori OB1a and 26 in Ori OB1b; as a result we 
confirmed 3 new members in Ori OB1a, one of which is substellar, 
and 19 new members in Ori OB1b, out of which 7 are at the substellar limit and 5 are
substellar. We looked into the presence of accretion signatures by measuring
the strength of the H$\alpha$ line in emission. Accordingly, we classified
the new members as having Classical T-Tauri star (CTTS) or Weak Lined T Tauri 
star-like (WTTS) nature.
We found that all the new members confirmed in Ori OB1a are WTTSs, 
while $39^{+25}_{-22}\%$ of the new members in Ori OB1b exhibit 
CTTS-like behavior, suggestive  
of ongoing accretion from a circum(sub)stellar disk. Additionally we found 
that none of the members confirmed in OB1a show near-IR color excess
while $38^{+26}_{-21}\%$ of OB1b members show H-K color excess.
These results are consistent with recent findings for low mass young stars 
in Orion OB1. The similarity in CTTS-like properties and near-IR excess across the substellar 
boundary gives support to the idea of a common formation mechanism for low mass stars and 
at least the most massive brown dwarfs. Finally, we remark the discovery of two new 
members classified as CTTSs, both exhibiting $\rm W(H\alpha) \la -140${ \AA }, 
suggesting significant ongoing accretion.

\end{abstract}
\keywords{Brown Dwarf: Very Low Mass Stars: Star Forming Regions:}

\section{Introduction}

One of the main goals of contemporary astrophysics is understanding
the processes of star and planet formation. In this context,
the formation of the least massive stars and brown dwarfs
is a key issue.
Because brown dwarfs (BDs) are objects with masses intermediate between
those of stars and planets 
\cite[$0.072M_\odot \ga M \ga 0.010 \> M_{\odot}$;][]{bar98,opp98},
understanding how they form can provide
important links between the origin of stars and planetary bodies.

Current observational efforts seek to
establish to what extent there is a continuity of the star
formation process across the substellar limit
\cite[$M\sim0.072M_\odot$;][]{bar98}, 
and wether the properties of the resulting
(sub)stellar populations depend on aspects like the surrounding
environment. 
Observational evidence shows that for some very
young populations the
kinematic and spatial distributions of very low-mass stars (VLMSs) and BDs
are similar, that there is continuity of the initial mass function
between VLMSs and BDs with masses down to $0.02M_\odot$
\citep{luhman07}, and that BDs accrete from
circum-substellar disks, much like low-mass stars do 
\cite[e.g.][]{jayawardhana02,muzerolle05}.
These results suggest
that the low mass star formation process extends across and below the substellar limit. 
However, these findings are based on observations of regions with 
ages between $1$ and $2$
Myr (Taurus, IC348, Trapezium and Chameleon I), much younger
than the $\sim 10$ Myr 
time span in which giant planets are expected to form \citep{cal05}.
Also, these studies do not
offer clear evidence on the dependency (if any) of the formation processes with 
the environment, because they compare different regions 
which do not share a common origin.

An important and required next step is the detection, minimizing spatial biases, 
of stellar and substellar pre-main sequence populations in star forming
regions that show a variety of
environmental conditions, span an age range from $\sim 1-10$ Myr,
and ideally share the same ``genetic'' pool.
Such samples can offer new insights into 
the formation and evolution of BDs:  a) The time dependence of 
indicators of disks and of disk accretion, over a span of 10 Myr, 
will offer the best view yet of
how disks evolve, under a diversity of environments and at both
sides of the substellar limit. b) The space distribution 
and kinematics of these populations will provide important constraints 
to the formation models, e.g. the dynamical ejection scenario 
\citep{reipurth01}, which suggests that BDs are stellar embryos 
ejected from their birth places as a consequence of dynamical 
interactions during the early stages of their formation, 
resulting in them being deprived from gaining 
mass from the surrounding gas.

The Orion star forming region meets all these requirements. 
It is a relatively nearby OB association ($\sim $ 400 pc; \cite{bri05}),
representative of the birthplace for the majority of stars in
our Galaxy \citep{bri07a},
containing regions with ages $\sim 1-10$ Myr and 
differing ambient conditions \citep{genzel89,bri05}. 
There is also a molecular cloud complex \citep{maddalena86},
in which the youngest star populations are embedded,
that shares the same kinematics as the older
stars \citep{bri07b}, suggesting that 
the stellar populations at various ages 
share a common origin. Throughout the whole region there is now a
significant number of confirmed pre-main sequence stellar members 
\citep{bri05,bri07b}; however, substellar objects have only been
confirmed so far in the younger, densest regions with small spatial extent,
such as the $\sigma$ Ori cluster \citep[e.g.][]{bejar99,bejar01,barrado03,caballero04,
sherry04,kenyon05,gonzalezgarcia06} 
or the Orion Nebula Cluster \citep{hill00,preibisch05,slesnick05}.
In order to look for the faintest, lowest mass members widely distributed across
the Orion OB1 association, down to and beyond the substellar limit, we are
conducting a large-scale optical-infrared 
search for BDs that spans $\sim $ 180 $deg^2$ over the entire region.
Our strategy is to combine deep, coadded, $R$-band and $I$-band data from our 
optical photometry obtained with the QUEST I CCD Mosaic Camera at Venezuela 
\citep{bri05}, with near infrared measurements from the Two Micron All Sky Survey
(2MASS) to select objects, which we call candidates, whose photometric
magnitudes and colors are consistent with those of BDs and VLMSs. 
We then obtain follow-up spectra to confirm membership 
and establish their stellar or substellar status. Finally, we characterize objects 
in terms of spectroscopic signatures such as strong $H\alpha$ emission, which
we use as a proxy for ongoing disk accretion.

In this work we present initial results in this effort to extend
the known young population of the entire Orion star-forming region
throughout the substellar limit. We selected optical/near infrared candidates 
within an area of $\sim $ 21 $deg^2$, encompassing part of the Orion OB1a 
and most of the Orion OB1b subassociations. We conducted
spectroscopic follow-up of candidates in an area of 0.8 $deg^2$ in OB1a,
and 2.5 $deg^2$ in OB1b.
The optical photometric observations are described in section 
\ref{photometry}. Candidate selection 
was done by defining regions in optical/near-IR color-magnitude and color-color
diagrams, as explained in section \ref{candidate_selection}. 
Section \ref{optical_spec} shows the results of the spectroscopic 
observations and spectral classification 
of a first sample of candidates, covering $\sim $3.3 $deg^2$. 
The criteria followed to establish membership and the stellar or substellar nature 
of the candidates are presented in section \ref{membership_bdnature}. 
In section \ref{tts_nature} we study the spectral types and $H\alpha$
emission in order to classify the new BDs and VLMSs
according to their Weak or Classical T-Tauri-like signatures,
and compare the fraction of both types of objects
with previous results obtained in the stellar low-mass regime 
($0.25 \la M/M_\odot \la 0.9 $) of the same region \citep{bri05,bri07b}. 
In section 7 we discuss the infrared emission properties of the new members,
and compare with similar mass objects confirmed as members in other subregions 
of Orion. In section \ref{extinctions_hrdiag} we estimate 
the extinctions, temperatures, luminosities and masses and plot the newly 
confirmed members in H-R diagrams. 
Finally in section \ref{summary_conclusions} 
we summarize the results and conclusions.

\section{Photometry}\label{photometry}

\subsection{Coadding of optical images}\label{coadd}

Since 1998 a multi-band (Johnson-Cousins BVRI) and multi-epoch large-scale survey 
($\sim $ 180 $deg^2$) of the Orion star-forming region 
($5h<\alpha<6h$,$-6\degr<\delta<6\degr$) is being performed with the J\"urgen 
Stock 1.0/1.5 Schmidt-type telescope and the QUEST-I camera, at the Venezuela
National Astronomical Observatory. The camera was developed by the QUEST collaboration
\citep{bal02} and consists of sixteen $2048 \times 2048$ pixel CCDs in a $4\times4$ 
array which, taking into account the gaps between CCDs, provides a field of view of
5.4 $deg^2$ with a scale of $\sim $ 1.02 $arcsec/pix$. The system is optimized for
observations performed in drift-scan mode at declinations near to the celestial
equator. In this mode of operation, 
the telescope is fixed while the sidereal motion is compensated 
by adjusting the angle of each row of CCDs, which ride on a common supporting
Invar rod, and by fine-tuning the readout frequency for each chip. 
Each CCD row is oriented in the north-south direction and is provided 
with a filter, resulting in a quasi-simultaneous 
observation in four different filters with an integration time per filter of $140 \sec$ 
at $\delta=0^o$. Thus, the system produces scans at a rate of $\sim 34.5 deg^2/hour/filter$.

Our main interest is the analysis of existing $I$-band observations,
for which the system attains a limiting magnitude $I_{lim} \sim 19.5$,
with a completeness limit of $I_{com} \sim 18.5$. 
In order to evaluate the faintest magnitude at which we could expect 
the substellar limit to correspond for each of the regions spanned by our
optical data, allowing for uncertainties in ages, distance and reddening,
we made the following estimates. Using the models of \cite{bar98}, the substellar limit
for OB1a on the 12.6 Myr isochrone is placed at $I\sim 18.1$ 
(for a distance modulus $(m-M)_{OB1a} = 7.59$ and 
extinction $A_V=0.6$; \citealt[]{bri05, bri07b}), and the
substellar limit for Ori OB1b on the 6.3 Myr isochrone is placed 
at $I\sim 18.1$ too (distance modulus $(m-M)_{OB1b} = 8.21$
and maximum extinction $A_V=2$; \citealt[]{bri05, bri07b}). 
We consider these isochrones as rough upper limits for the age
of each subregion based on the age ranges reported by \cite[]{bri05}
for OB1a and OB1b. 
Because the substellar limit in each region is near the completeness limit
of our individual I-band observations, we use a coadding technique in order
to increase the signal to noise ratio (SNR) and therefore the limiting and
completeness magnitudes of our optical data.

For the coadd we selected 8 individual (single) scans centered at $\delta = -1^o$, 
that were obtained in the $R$ and $I$ bands between December 1998  
and January 1999. The corresponding 
observation log is shown in Table \ref{table:questlog}\footnote{This publication makes use of 
the Sistema de Colecci\'on de Datos Observacionales del Telescopio Schmidt version 2000 
\cite[SCDObs2000;][]{ponsot07}}. The sum was performed using the packages 
\texttt{Offline} \citep{bal02} and \texttt{DQ} 
developed by the QUEST collaboration. 
\texttt{Offline} automatically processes an individual scan considering 
each CCD of the array as an independent 
device, reducing the images by bias, dark current and flat field, using
calibration images obtained in drift-scan mode during the observation
runs (see \cite{bal02} for further details on the reduction process).
After these corrections, \texttt{Offline} performs the detection 
of point sources, aperture photometry, and finally computes 
the astrometric matrix for each CCD image, based on the USNOA-2.0 catalog \citep{mon98}. 
\texttt{DQ} processes the raw scans, regarding each CCD from different scans as 
independent devices; then does the bias subtraction, dark current and flat field
corrections, following the same procedure as employed by \texttt{Offline}.
Using the astrometric matrices produced by \texttt{Offline}, the \texttt{DQ} software
computes the offsets, rotations and scale differences between images of the same
area of the sky obtained from different scans. With these corrections
\texttt{DQ} produces, for each scan, a new output ``scan'' composed by a 
set of processed images that can be added pixel by pixel to produce 
the final coadded scan. The final gain in depth was $\sim 1$ magnitude 
and the resulting coadded scan 
encompasses the region $75\degr<\alpha<85\degr$,  $-2.2\degr<\delta<0.1\degr$ 
for the $R$ and $I$ filters. 

\begin{figure}
\includegraphics[angle=270,scale=0.70]{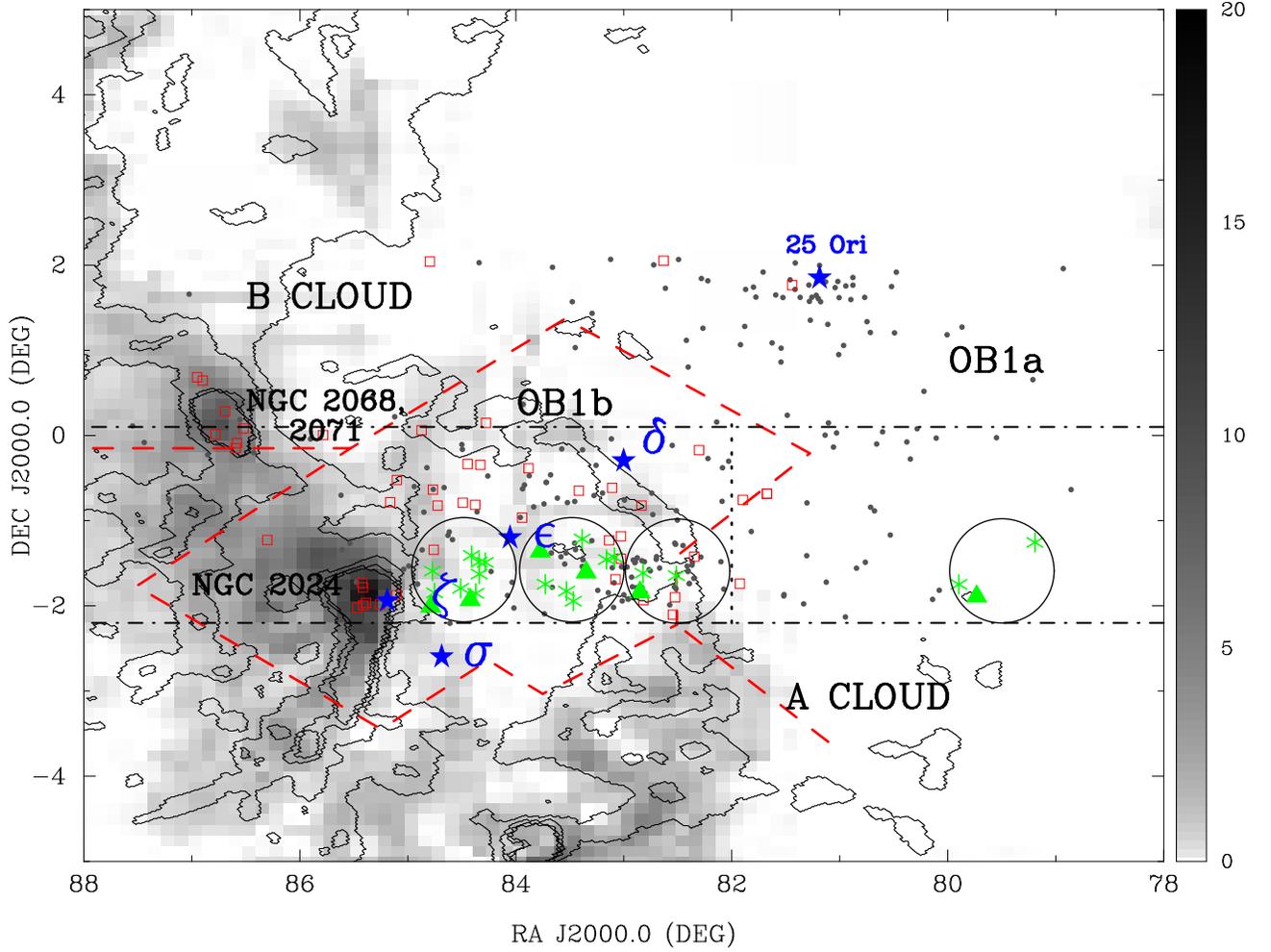}
\caption{Spatial location of the newly discovered VLMSs and BDs. Triangles
and stars indicate BDs and VLMSs respectively, spectroscopically confirmed as members
within the Hectospec fields (solid line circles). Members classified by \cite{bri05,bri07b}
as CTTSs are indicated with small empty squares and WTTSs with filled dots. 
The horizontal dot-dashed lines indicate the north and south limits of our scans centered
at $\delta_{J2000.0}=-1.0\arcdeg$. The dashed line region marks the boundary of the OB1b
and OB1a regions as defined by \cite{wh77}. The dotted vertical line indicates 
the adopted spatial limit between OB1a and 
OB1b subregions into the strip of our photometric sample as we explained 
in section \ref{coadd}. The gray scale map shows the integrated $^{13}$CO emissivity
from \cite{bally87}.
The isocontours correspond to the dust extinction map of 
\cite{schlegel98}, in $E(B-V)$ steps of 0.1 to 1.5 mag.
}\label{spatial_plot}
\end{figure}

Our first step in defining the photometric candidate sample is the
selection in I vs I-J diagrams of objects located above the isochrone that
defines the upper age limit for each subassociation, because it is in
these regions of the color-magnitude diagrams that the pre-main sequence
populations are expected to fall. Because of the differences in age and
distance we need to
apply this procedure independently for each subassociation,
which requires the definition of a spatial limit between the two.
Using the integrated $^{13}CO$ emissivity observed in the region by
\cite{bally87}, and the isocontours corresponding to the dust extinction map of
\cite{schlegel98}, as indicators of the position of the molecular clouds
associated with OB1b, we considered $\alpha \sim 82\deg$ 
as the rough limit between OB1a and OB1b
within the 2.3 deg wide strip spanning the photometric sample analyzed in this work.
This spatial limit between the two subassociations is consistent with that
discussed by \cite{bri05}, who consider the molecular ringlike structure 
roughly centered on the Orion Belt star $\epsilon$ Ori, as a 
good tracer of the extent of the OB1b subassociation; within this
gas ring are found the majority of low-mass PMS stars 
classified by them as members of OB1b.
Figure \ref{spatial_plot}
shows part of the entire area of our Orion optical survey, as well as the
strip centered at $\delta_{J2000.0}=-1\arcdeg$ considered here, the dust 
extinction map and $^{13}CO$ emissivity used to define the limit between 
both regions and the fields of the spectroscopic followup analyzed in this work.

\subsection{Optical photometry and 2MASS data }\label{optical_2mass_phot}

The detection of point sources in the coadded scan was performed with the 
\texttt{daofind} task in IRAF\footnote{IRAF is distributed by the 
National Optical Astronomy Observatories, which are operated by the Association
of Universities for Research in Astronomy, Inc., under cooperative agreement 
with the National Science Foundation.}, extracting objects with $R$ and $I$-band 
fluxes above $3\sigma_{sky}$. The small amount of crowding in our images 
($\sim$ 1.4 sources/arcmin$^2$) justifies the use of aperture photometry. 
We used the \texttt{phot} task in IRAF, adopting an
aperture radius (computed from growth curves) of $4\arcsec$, which is $\sim 1.4$ 
times the typical FWHM in both bands. Because our pixel scale is $1 \arcsec$ and
the typical FWHM of the images is $2.5\arcsec - 3 \arcsec$, each stellar profile is well sampled.

The astrometric solutions were computed using the WCStools package \citep{min99}.  
and astrometric standards of the Guide Star Catalog GSCII\footnote{The Guide 
Star Catalog was produced at the Space Telescope Science Institute under U.S. 
Government grant. These data are based on photographic data obtained using the 
Oschin Schmidt Telescope on Palomar Mountain and the UK Schmidt Telescope.}, 
yielding a median offset between $R$-band and $I$-band catalogs of 
$0.89\arcsec$ with a standard deviation of $0.78\arcsec$.

The photometric calibration to the Cousins system was performed by defining 
a set of 416 \emph{secondary photometric standards}, 
located in four fields within our $\delta_{J2000.0}=-1\arcdeg$ Orion strip.
We made sure that each of the 4 strips at constant declination that compose
a single driftscan observation, as produced by each of the 4 rows of detectors
in the QUEST-I camera, would enclose roughly 100 of these secondary
standard stars.
The stars were
chosen as objects known not to be variable from our multi-epoch survey data;
only those flagged as non-variable at the $\le 0.05$ mag level were used.
The secondary standards span a range of brightness 
but are not saturated in any single
observation, and they are not too faint,
so that each star would yield a S/N ratio $\ga 50$. 
The secondary standards were in turn calibrated using 
12 primary photometric standard stars located in the 
SA92 and PG0231 Landolt fields \citep{lan92}, observed
at various airmasses under photometric conditions, 
with the 4-SHOOTER CCD camera
on the 1.2 m telescope at the Smithsonian Astrophysical Observatory.
The seeing was $\sim 2$\arcsec{ } throughout the observations.
The 4-SHOOTER camera contains four $2048\times 2048$ Loral CCDs separated
by {45\arcsec } and arranged in a $2\times 2$ grid. After binning $2\times 2$
during readout, the plate scale was ${\rm 0.67\arcsec \> pixel^{-1}}$.
In order to achieve more uniform and consistent measurements, we
placed all the standard stars on the same CCD detector (chip 3).
Bias and flat field corrections were applied to the raw images using
the standard tasks in the IRAF \texttt{ccdproc} package. We then
performed aperture photometry with the \texttt{apphot} package,
and derived zero points, extinction coefficients and color terms
using the \texttt{photcal} package.
The photometric errors of the secondary photometric standard stars are
$\sigma  < 0.05 $ in the R and I bands. 

The calibration of sources from the coadded images was performed independently
for each CCD using standard IRAF tasks, yielding an $RMS= 0.037$ mag and $0.042$ mag 
in the $R$ and $I$ bands respectively. Sources without detections
in the $R$ band were calibrated independently in the $I$ band using zero points only. 
Therefore, we obtained two sets of data calibrated in different ways: one including
an R-I color term and the other without it. In the transformation equation 
used to convert instrumental magnitudes to the secondary standard star system,
the R-I color coefficient has a rather small mean value of $-0.046$ for the I band. 
The observed R-I color of the candidates are  in the interval $1.24<R-I<2.4$, 
therefore, the difference between I magnitudes computed using a color term 
and those computed without it is $0.05$ magnitudes for the bluest and brighter 
candidates (expected to have an $\sim M3$ spectral type) and 
$0.11$ magnitudes for the reddest an fainter candidates (expected to have an
$\sim M7$ spectral type). The final optical data catalog includes 
$\sim 223000$ objects with typical instrumental errors of $0.04<\sigma_I \le 0.2$ at 
$19 \ge I \ge I_{lim}$ and $0.05<\sigma_R \le 0.2$ at $20 \ge R \ge R_{lim}$.
Therefore, the calibration using secondary standard stars offers adequate photometric 
accuracy for the VLMSs and substellar candidate selection and 
the subsequent analysis presented at this study.
The $3 \sigma$ limiting magnitudes obtained are $R_{lim} = 21.5$ and $I_{lim} = 20.7$,
with completeness limits of $R_{com} = 20.3$ and $I_{com} = 19.0$. 
We define here completeness 
as the value at which the magnitude distribution departs from a linear behavior. 
Saturation occurs at $R_{sat} \sim I_{sat} \sim 13.5$.

A total number of $\sim  127000$ sources were observed in our survey region by 
the 2MASS
project, with 
limiting magnitudes of $J_{lim}=18.9$, $H_{lim}=18$, $K_{lim}=17.4$ and 
completeness limits of $J_{com}=16.6$, $H_{com}=16$, $K_{com}=15.4$, computed following
the same procedure used for our optical data. Typical uncertainties in the 2MASS data
are $0.1<\sigma_{J,H,K} < 0.5$ between completeness and limiting magnitudes for the three bands.
A total number of $\sim  106000$ objects from the optical catalog were identified 
in the 2MASS data, resulting in a median offset
between both catalogs of $1.41\arcsec$ with a standard deviation of $0.89\arcsec$.

\section{Candidate selection}\label{candidate_selection}

The candidate selection was performed using optical-infrared 
color-magnitude and color-color diagrams. This method has proved 
successful in identifying VLMS and BD candidates in young clusters and various 
star-forming regions \citep{bri02,luh03a}.

\begin{figure}
\includegraphics[scale=0.70]{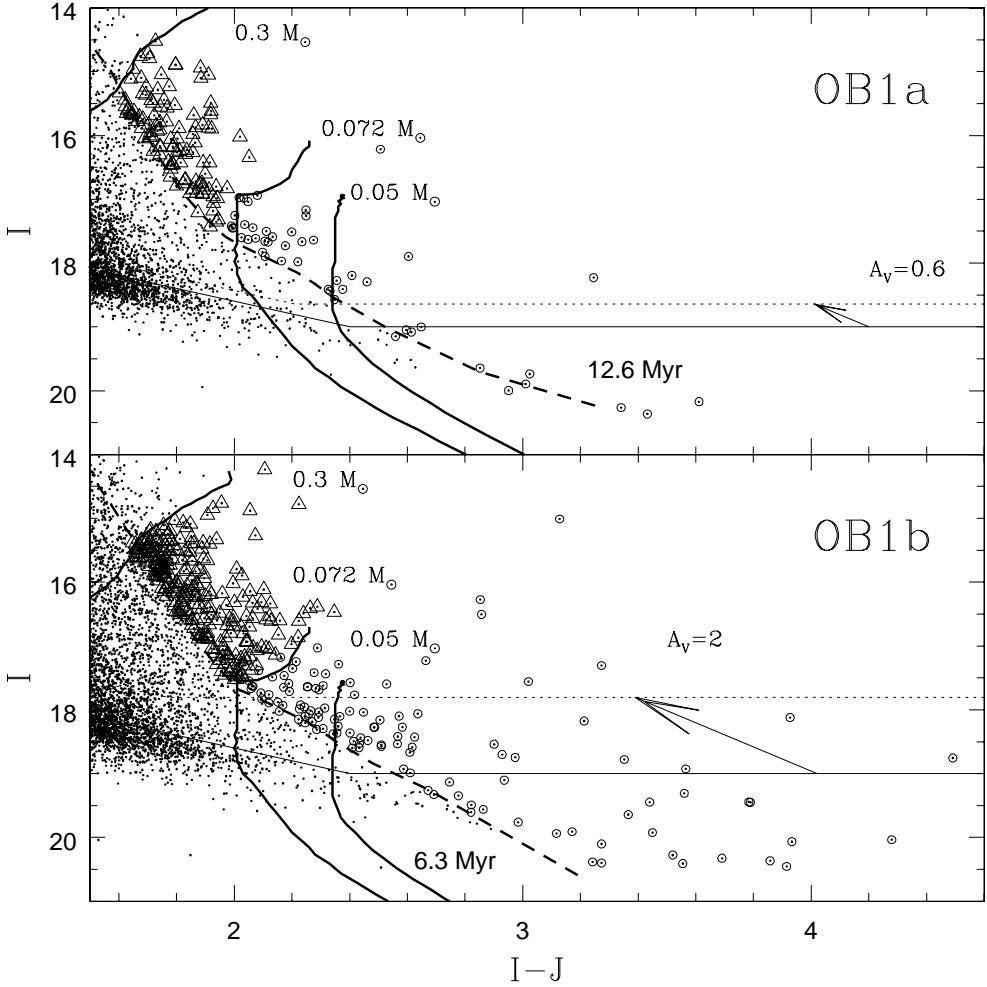}
\caption{First selection of candidates. 
Dots represent field sources located under the isochrone 
that defines the upper age limit of each region, based on results
from \cite{bri05} (12.6 Myr for OB1a and 6.3 Myr for OB1b; dashed lines), or PMS
candidates located above the $0.3M_{\odot}$ 
evolutionary track that were rejected as candidates. 
Sources above the isochrones, between the $0.3M_{\odot}$ and $0.072M_{\odot}$ 
evolutionary  tracks (open triangles), 
and below the substellar limit ($0.072M_{\odot}$ evolutionary track; open circles), 
were selected as initial candidates. 
The thin solid lines represent the completeness of the 
photometric sample, and dotted lines show the completeness assuming 
$A_V=0.6$ for OB1a,
and $A_V=2$ for OB1b. 
These extinctions correspond to the representative upper values 
measured in our spectroscopic sample for each 
region, resulting in a mass completeness of 
$\sim0.05M_{\odot}$ for OB1a and $\sim0.072M_{\odot}$ for OB1b. 
We used models from \cite{bar98} and
a distance modulus of $7.59$ for OB1a,
and $8.21$ for Ob1b \citep{bri05}.
The reddening vectors were computed using the \cite{car89} extinction law with $R_V=3.1$,
and temperature-spectral type relationship from \cite{luh03a}.}\label{IvsIJ}
\end{figure}

\begin{figure}
\plotone{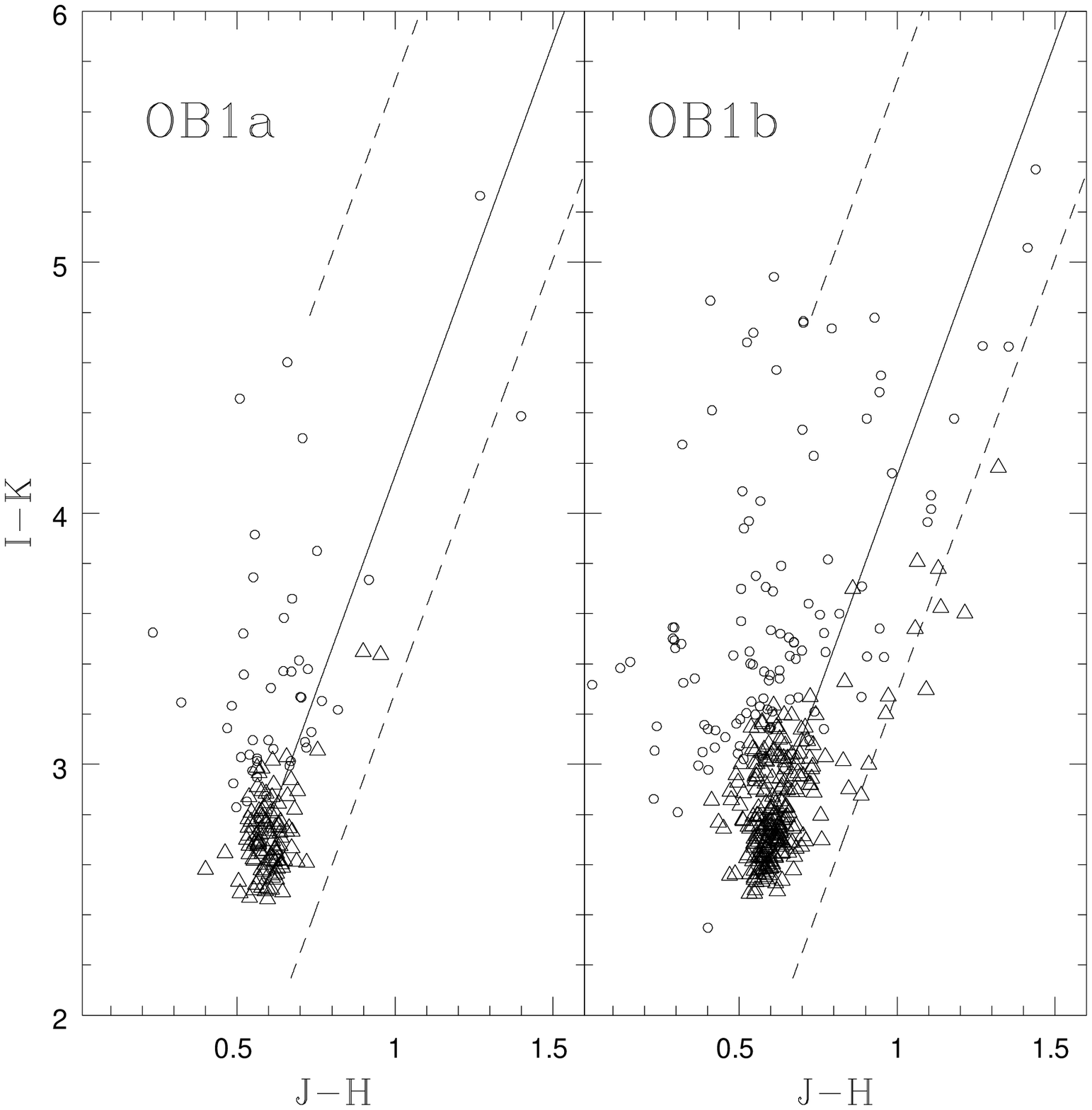}
\caption{Final step in the selection of candidates. 
The symbols are the same as in Figure \ref{IvsIJ}. Solid lines 
represent the reddening line for M5 objects and dashed lines the reddening lines
for M1 (lower) and M9 (upper) objects,
computed using the \cite{car89} extinction law with $R_V=3.1$, \cite{bar98} models, 
and temperature-spectral type relationships from \cite{luh03a}. 
On average, sources falling above the M5 reddening line are later than M5 at different $A_V$ values
and were selected as final candidates.}\label{IKvJH}
\end{figure}

We compared the optical and infrared magnitudes and colors
with isochrones and evolutionary tracks
from \cite{bar98} models, in
diagrams that included $I$, $J$, $H$ and $K$-band data. 
As mentioned in section \ref{coadd},
we assumed a distance modulus of $(m-M)_{OB1a} = 7.59$ for OB1a, and 
$(m-M)_{OB1b} = 8.21$ for OB1b \citep{bri05}. 
The first step was the selection in the $I$ vs $I-J$ diagram of objects 
placed above isochrones assumed to be a reasonable
upper age limit for each subassociation, 12.6 Myr for OB1a and 6.3 Myr for OB1b
(\S\ref{coadd}). 
In addition, because we are interested here in the lowest mass PMS stars
and BDs, we required 
that candidates were located below 
the $0.3 \> M_{\odot}$ evolutionary track of the \cite{bar98} models\footnote{The
census and study of the Orion OB1 PMS stars with masses above 
$0.3 \> M_{\odot}$ is the subject of works like \cite{briceno01,bri05}.}. 
These diagrams are shown in Figure \ref{IvsIJ}.
The objects selected from the $I$ vs $I-J$ diagrams were then plotted on the $H$ vs. $I-K$ 
plane. There we checked their positions with respect to the same isochrones and evolutionary 
tracks as we used in the $I$ vs $I-J$ diagrams, in order to verify the consistency 
of the first selection step. Finally, in order to  reduce the contamination 
from reddened early type (and therefore more massive) stars,
we selected only objects located above the 
reddening line corresponding to a spectral type M5 
in the $I-K$ vs $J-H$ diagrams shown in Figure \ref{IKvJH}.
On average, only objects later than M5 ($M<0.1 M_{\odot}$), 
at different reddening values, are expected in this region of the diagram
as we show in figure 3.

The reddening lines and vectors were computed according to the \cite{car89} extinction
law with $R_V=3.1$ (representative of OB associations), \cite{bar98} models, 
and temperature-spectral  type relationships from \cite{luh03b}. The resulting 
catalog includes 64 VLMSs candidates
in OB1a and 118 in OB1b, and 58 BDs candidates in OB1a and 77 in OB1b.
Based on the completeness limits in the $I$ and $J$ bands, and
\cite{bar98} models, we estimate the photometric sample to be complete down to 
members with masses $0.05\> M_{\odot}$ with $A_V \le 0.6$ for OB1a, and 
masses $0.072\> M_{\odot}$ for OB1b with $A_V \le 2$ as we show in Figure \ref{IvsIJ}. 
Both reddening values are representative of the typical upper limits measured on our 
spectroscopic sample (see below) and in low mass stars confirmed in both
subregions by \cite{bri05,bri07a}.

\section{Optical spectroscopy}\label{optical_spec}

\subsection{Spectroscopic observations}\label{spec_obs}

Though the photometric candidate selection technique is particularly 
sensitive to young VLMSs and BDs, a fraction of $\sim 25-30$\% 
of the sample is expected to be composed of background and foreground objects, 
contaminating the candidate lists. Actually, the field dwarfs 
2MASS 05395200-0059019 \cite[L5V;][]{fan00} 
and the  2MASS 05012406-0010452 (L4V; Reid et al. in preparation)  
were detected in our photometric survey with $R=20.27\pm 0.16$, 
$I=17.32\pm 0.01$ ($R-I = 2.95$) and $R=20.14\pm0.08$, $I=18.23\pm0.02$ 
($R-I = 1.91$) respectively and are included in our candidate list.
Also, background late-type giants may be present.
Therefore, spectroscopic confirmation of membership to the association
is necessary.

We obtained low resolution spectra of 
a first sample composed of 30 objects: 4 candidates located in OB1a
and 26 in OB1b. 
We used the Hectospec Multi-fiber spectrograph on the 6.5 MMT telescope 
\citep{fab98}. In its $f/5$ configuration the telescope 
offers a $\sim 0.8\deg^2$ field of view, in which 300 fibers of the spectrograph 
can be placed. We used the 270 $groove \> mm^{-1}$ grating, that provides 
a spectral resolution of $6.2${ \AA} with a spectral coverage from 3700 to 9150{ \AA}. 
Each fiber subtends $1.5\arcsec$ on the sky and requires 
accurate coordinates for positioning. We used 2MASS coordinates for 
all our candidates. 

The observations were performed as part of the spectroscopic followup of
the CIDA Variability Survey of Orion 
\citep{bri05} in which the main goal is the identification of low mass stars
in the range $0.2 < M/M_{\odot} < 0.9$. Because of this, the spatial coverage 
in each OB1a and OB1b was dictated by the number of fields observed with Hectospec
in each region for the shallow survey, out of which 
only fields containing VLMSs and BDs candidates from our deep coadd are considered here. 
The 4 candidates located in OB1a all fall within one Hectospec field, spanning 
$\sim 0.8\deg^2$. The 26 objects in OB1b are distributed over a larger area
of $\sim 2.5\deg^2$ (three Hectospec fields). 
For all fields the total integration time was $3\times900 \sec$. 
Because in Hectospec all objects placed in the same field have 
the same integration time, many of our faint VLMSs and BDs candidates had
low $S/N$. However, this did not affect in a significant way
the spectral classification as we explain in the following section.

All the spectra were processed, extracted and wavelength calibrated by 
S. Tokarz at the CfA Telescope Data Center, using customized IRAF routines 
and scripts developed by the Hectospec team. A typical number of 37 
fibers per field were used to obtain sky spectra; these were combined 
and the resulting spectra subtracted from the target spectra.
The fields in the OB1b subassociation were selected far enough from the
the molecular clouds to avoid problems due to the high sky background
caused by nebulosity, however, some objects placed in the easternmost OB1b field
do show some contamination by cloud emission lines (however, these lines
did not affect the spectral classification).
The spectra were wavelength calibrated using arc lamp data. The appropriate 
correction for the sensitivity function of the detector was applied, and the 
effects of the atmospheric extinction were computed, using standard 
IRAF routines based on observations of the spectrophotometric standard star
BD+284211. Light sources inside the fiber positioner of Hectospec contaminate
$\sim  1/3$ of the fibers at the red edge, starting at $\sim 8500$ {\AA }
(N. Caldwell, private communication), therefore, during the analysis we ignored 
data at wavelengths beyond this value. The observations log is shown in
Table \ref{table:HECTOSPEClog}.

\subsection{Spectral classification}\label{spec_clasif}

We performed the spectral classification based on the comparison with
of temperature-sensitive spectral indices. We extended the semi-automated 
procedures of \cite{her04} and \cite{sic05} for their application 
to VLMSs and BDs, by adding indices measured for the 16 spectral features
listed in Table \ref{table:molecfeatures}, and a library of standard 
spectra of field dwarfs from \cite{kir99}.
Our set of spectral indices span a wavelength range 4775-7940{\AA }, 
including spectral features such as TiO absorption bands, 
that increase in strength from K5 to M7 spectral types, and VO absorption bands, 
that start dominating spectra from M7 to early L types.

\begin{figure}
\plotone{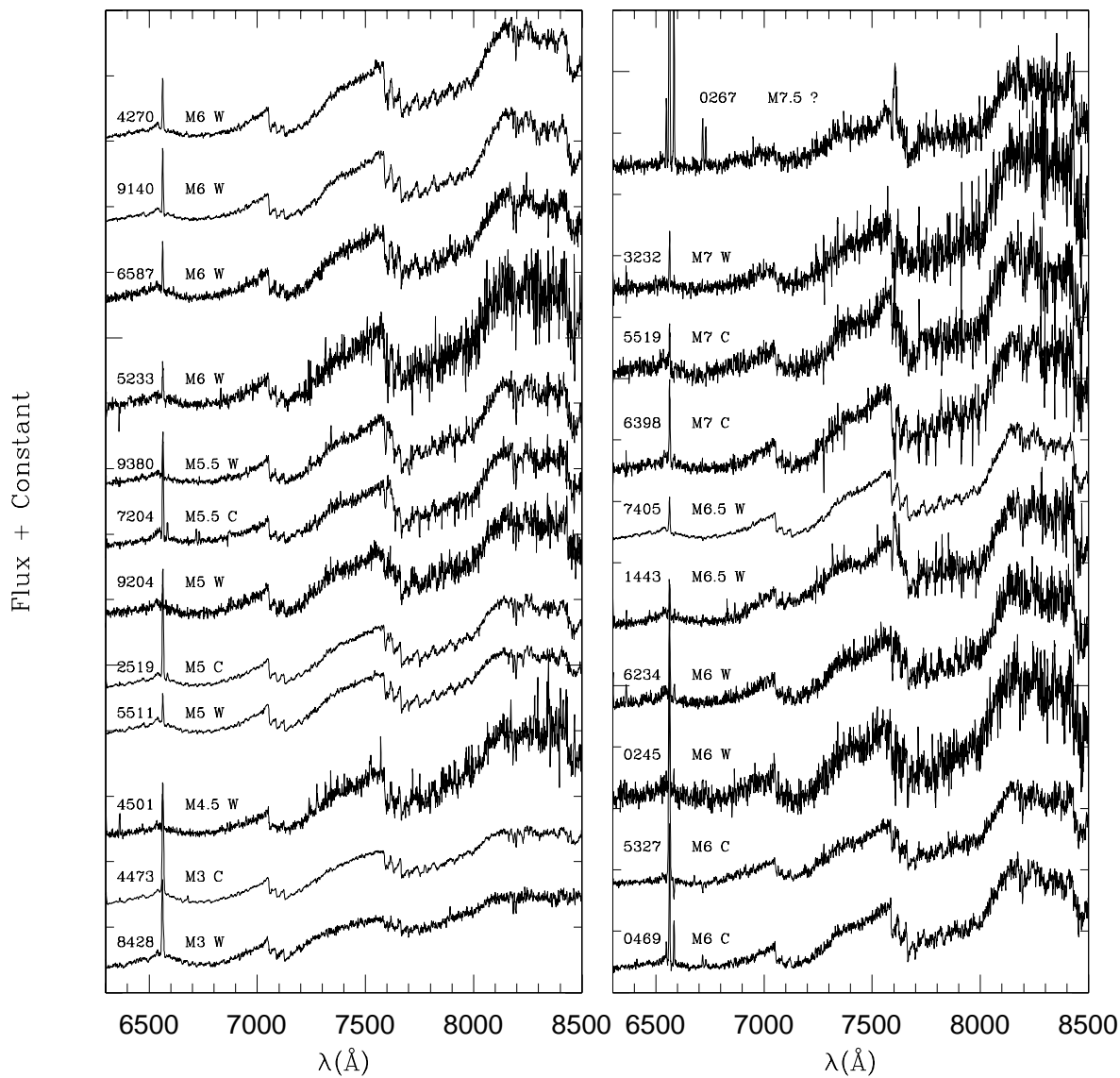}
\caption{New members confirmed in this work. Each spectrum shows the last four 
characters of its 2MASS ID, and the spectral type followed by its classification 
as CTTSs or WTTSs. All spectra were normalized at $7500$ \AA \space 
and are shown at their original resolution of $6$ \AA. Member 0267 does not have a clear 
classification as CTTS or WTTS, as is explained in subsection 6.1.}\label{Mspectra}
\end{figure}

Spectral types for noisy spectra were checked directly
with the standard star spectral sequence using the visual extinction 
as a free parameter to get a more accurate result. The classification \footnote{This 
procedure has been done using the code SPTCLASS, available at 
http://www.astro.lsa.umich.edu/$\sim$hernandj/SPTclass/sptclass.html} 
is shown in Table \ref{table:data}, and Figure \ref{Mspectra} shows the 
spectra, with adopted spectral types of the new members shown in labels.

\section{Association membership and substellar nature}\label{membership_bdnature}

\subsection{Membership diagnostics}\label{membership}

Candidates classified as late-M type objects can be divided into three different 
groups: first, the VLMSs and BDs that belong to the Orion OB1 association; 
second, background stars composed by late type giant stars; and third,
foreground,  old M-dwarf field stars. 
We have used several criteria to assign Orion OB1 membership to each candidate. 
These include the location above the ZAMS in color-magnitude diagrams,
$A_V$ values close to the characteristic absorption towards each region,
and spectroscopic indicators that characterize young, low-mass
objects.

\begin{figure}
\plotone{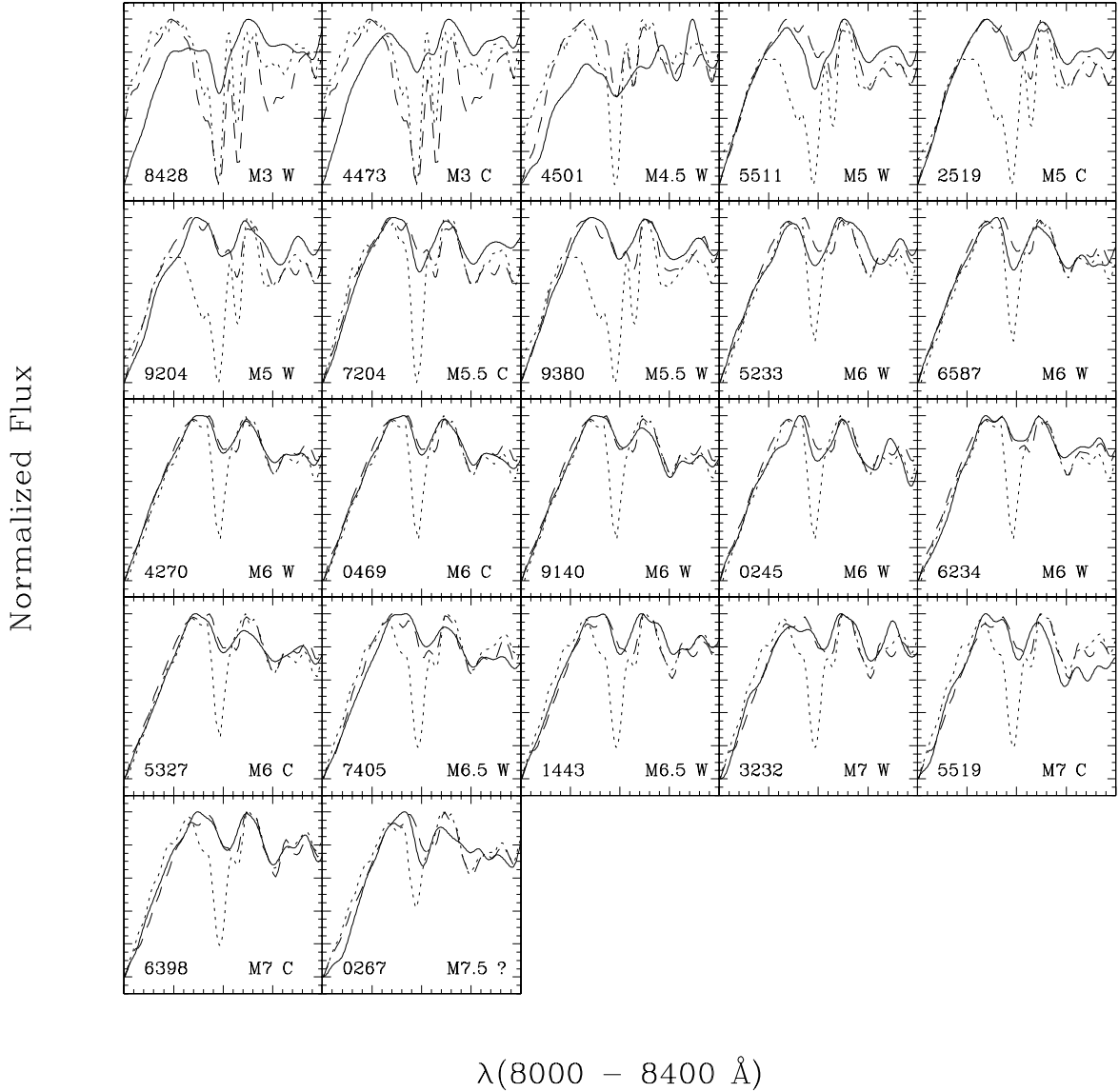}
\caption{The Na I $\lambda 8195$ absorption lines 
in the newly confirmed members (solid line), 
for field dwarfs (dotted line; \cite{kir99}),
and for young VLMSs and BDs from Chamaeleon I region
($\sim$ $2Myr$; dashed line; \cite{luh04}). 
New members and comparison templates have the same spectral 
types. All spectra were smoothed to a resolution of 
$16$ {\AA} and normalized in the interval 8000-8400\AA.}\label{na_lines}
\end{figure}

First, we consider the H$\alpha$ $\lambda 6563$ line in emission,
with equivalent width increasing toward later spectral types \citep{bar03,whi03}.
Strong H$\alpha$ emission is a common feature of chromospherically active young objects, 
and of young objects accreting from a circumstellar disk 
\citep[for the largest equivalent widths; e.g.][]{muzerolle05}.
By adopting this criterion we eliminated contamination from background giant stars.
Still, H$\alpha$ emission can occur in dMe stars. In order to 
distinguish PMS objects from these late type field stars,
and considering that PMS objects are still contracting, 
we chose as an additional youth indicator the strength of the Na I $\lambda 8195$ 
absorption feature. This doublet  
is sensitive to surface gravity and varies significantly 
between dwarf field stars and PMS objects;
equivalent widths in young M dwarfs fall between those of field 
M dwarfs and M giant stars \citep{luh03b}. 
Figure \ref{na_lines} shows the Na I lines of M type objects confirmed as members,
superimposed on spectra of field dwarfs from \cite{kir99}, and of young M dwarfs 
from \cite{luh03b} of the same spectral type. 
All these spectra were smoothed to the same resolution of 16{ \AA }, and normalized
in the range 8000-8400{ \AA}. Both criteria used simultaneously offer
a good membership indication. Additionally, the Av values obtained 
in this sample are in good agreement with
the characteristic values measured in both subassociations by
\cite{bri07b} for low mass stars.
Summarizing, M-type objects were classified as members
if they showed H$\alpha$ emission, Na I absorption clearly 
weaker than in field M dwarfs, and $A_V$ consistent 
with the known values for each subassociation. 
The confirmed new members (3 in OB1a and 19 in OB1b) are shown in Table 
\ref{table:data}.
In the OB1a candidate sample we identified one late type (M6) object 
from the field, and in OB1b we classified 7 objects as field stars,
4 with late spectral types (M3, M5, M6 and M7) and 3 reddened early type
stars. 

Taking our entire spectroscopic sample of 30 objects, we find
that 8 were classified as field stars (1 out of 4 in OB1a and 
7 out of 26 in OB1b). Applying the exact test for the success rate
in a binomial experiment (R-Statistical Software, \citealt{ihaka_gentleman96}),
we find that the proportion of objects classified correctly
as members is 0.73, with a 95\% confidence interval of 0.54-0.87.
Therefore, the selection technique described here identifies
correctly 73.3\% of objects.

\subsection{Substellar status}\label{bd_status}

According to the models of \cite{bar98}, objects with masses at the substellar
limit ($0.072\> M_{\odot}$) and ages between 7.9 Myr and 3.2 Myr (that correspond
to the mean ages of OB1a and OB1b computed by \cite{bri05} based on
PMS low-mass stellar populations), have effective temperatures
of 2936 K and 2993 K respectively. Therefore, adopting the temperature to spectral type
relationship from \cite{luh03b} the substellar limit at these ages corresponds
to a spectral type between M6 (2990 K) and M6.5 (2935 K). Following this criterion, 
we consider as BDs those members with spectral types
later than M6. In the Hectospec field located in the OB1a subassociation we confirmed 
1 VLMS with spectral type M4.5, 1 VLMS at the substellar limit with spectral type M6, 
and 1 BD with a spectral type M7.
In the Ori OB1b subassociation we identified 14 VLMSs, out of which 7 have M6 spectral 
types, and 5 BDs (two with spectral types M6.5, two M7 and one M7.5).

\section{T Tauri signatures across the substellar limit}\label{tts_nature}

The paradigm for low-mass star formation draws a picture in which
a central low-mass star (T Tauri star)
accretes mass from a circumstellar gas and dust disk,
via magnetospheric accretion. Meanwhile the inner regions of
the circumstellar disk are disrupted by the stellar magnetosphere,
which then channels viscously accreting material out of the disk plane and
towards the star along magnetic field lines \citep{konigl91,shu94}.
Signposts of this process,
which have been directly observed in CTTSs, include:
 a) infrared emission from dust at a range of temperatures in the disk
heated by stellar irradiation and viscous dissipation \citep[from $\sim 1500^o$K 
at a few stellar radii, down to $\la 50^o$ K at $\sim 100$ AU, e.g.][]{mey97,har98,muzerolle03};
 b) blue/UV continuum excess
emission from the accretion shock formed as accreting material falls
onto the stellar surface \citep{valenti93,hartigan95,calvet_gullbring98};
 c) broad permitted emission lines produced in
the ballistic magnetospheric gas flows \citep{muzerolle01}; and d)
forbidden emission lines produced in accretion-powered winds
and jets \citep{hartigan95}.

Recent studies have searched for similar characteristics in the lowest-mass
CTTSs and young BDs, as one way to ascertain whether
similar formation mechanisms apply between both types of objects. 
There is indeed considerable evidence for magnetically-mediated 
disk accretion, including broad permitted emission line profiles 
(especially H$\alpha$) and optical
continuum veiling in BDs \citep[e.g.][]{whi03,muzerolle05,jay03}.  
Furthermore, the infrared excess emission detected in many objects 
confirms the presence of irradiated circumstellar disks 
\citep{muench01,liu03,jay03,mohanty04} and high resolution
observations of H$\alpha$ emission allowed the measurements of the mass accretion 
rates in BDs \citep{muzerolle00} resulting in a much lower value 
than stellar-mass CTTSs with similar ages \citep[e.g.][]{muzerolle05,luhman07}.
It now seems well established that 
there is a continuum of CTTS-like properties across the substellar limit,
shared by both the lowest-mass CTTSs and at least the higher mass young BDs.

Low-resolution spectra provide an appropriate means to look for H$\alpha$
emission in CTTS. The dividing line between accreting 
(CTTSs) and non-accreting  (WTTSs) was first 
set in terms of  H$\alpha$ emission at W(H$\alpha$) $=-10$ {\AA} by \cite{hbc88}. 
More recently, \cite{whi03} have revised this classification. In their new
scheme, a star is considered to be a CTTS if $\rm W(H\alpha) \le -3$ {\AA} in the range K0-K5,
$\rm W(H\alpha) \le -10$ {\AA} for K7-M2.5,  $\rm W(H\alpha) \le -20$ {\AA} for M3-M5.5,
and  $\rm W(H\alpha) \le -40$ {\AA} for M6-M7.5.
Here we followed their criteria to classify an object as a CTTS or a WTTS.
Additionally Figure 6 shows the $HeI$ emission
for the spectra in which this line was detected. Figure \ref{wha_ij} shows 
the classification, with $\rm W(H\alpha)$ plotted 
as a function of the spectral type, with the 
CTTS/WTTS divide from \cite{whi03} indicated by the dotted line. The results are 
summarized in Table \ref{table:data}.

\section{Disks among the widely spread PMS population of Orion OB1}\label{disks}
 
\subsection{Trends in accretion-related indicators}\label{TTSfraction}

The fraction of objects exhibiting CTTS-like properties can be used
as an indicator of the population of disk-bearing objects which are
still accreting from circum(sub)stellar disks. How these numbers compare
among samples with differing ages, but presumably sharing a common origin,
can provide insight into the time scale for the
shutting off of accretion near and below the substellar boundary.

Using the same disk indicators we consider here, the first 
determination of the CTTS and WTTS fractions for the low-mass,
widespread stellar population in Orion by \cite{bri05}, over an area
of $\sim 68\> deg^2$ showed that 11\% of the
members in OB1a and 23\% of the members on OB1b are CTTS. 
More recently \cite{bri07b} found a lower CTTS fraction of 
6\% in Ori OB1a and 13\% in Ori OB1b. However, this study was
restricted to a much smaller area (totaling $\sim 1.6\> deg^2$).
It is worth noting that because there are regions with higher concentrations 
of CTTS, like the very young clusters NGC 2024/2068 located on the
Orion molecular clouds (e.g. see Figures 5 and 6 of \citealt{bri05}),
the CTTS fraction on small spatial scales is strongly dependent
on the location of the region being considered.

The very small size of our sample in OB1a (only 3 members) does
not allow us to derive any meaningful estimate, other than saying
that it is completely dominated by WTTSs. In OB1b, we find that 7 of the
18 members that could be assigned a TTS type, are CTTS
(because of suspected contamination from nebular emission, 
05390910-0200267 could not be classified as either WTTS or CTTS). 
If we again
apply the exact test for the success rate in a binomial experiment 
(R-Statistical Software, \citealt{ihaka_gentleman96}),
we find that the proportion of CTTS in OB1b is 0.39, with a 95\% confidence 
interval of 0.17 - 0.64; the large size of the confidence interval
being due to the relatively small sample size.
These results are largely consistent with previous findings,
in the sense that the OB1a population is dominated by non-accreting
PMS stars, whereas the accretor fraction in the younger OB1b 
region can be estimated in the range $\sim 13 - 40$\% depending
on the region being considered and the sample size.

Our findings suggest that the overall number of accretors
among VLMSs and BDs in OB1b is similar to that determined for low-mass
($M \ga 0.3 M_{\odot}$) stars by \cite{bri05}. Also, it seems that, as it happens in 
the low-mass PMS members, the VLMS/BD accretor fraction
falls off by a significant amount once the
population reaches ages $\sim 8-10$ Myr. Determining the detailed
evolution of the accretion fractions is similar between
low-mass stars and VLMSs/BDs will require larger samples.

\subsection{Near IR disk indicators}\label{ir_excess}

We analyzed the $J$, $H$ and $K$ band emission of all the new members confirmed
in this work. Figure \ref{JHvsHKOB1} shows the observational $J-H$ vs. $H-K$ 
diagrams with the members classified as  WTTSs and CTTSs. The members 
confirmed in OB1a (age $\sim8$ Myr) do not show excess emission in either color.
In the $\sim4$ Myr old OB1b, where a fraction of the low-mass PMS stars ($M>0.3M_\odot$) 
identified by \cite{bri05} show infrared excesses, 
7/19  members confirmed in this work 
(1 BD WTTS, 4 VLMSs WTTSs and 2 VLMSs CTTSs) have $H-K>0.45$ and lie to the right 
of the reddening line for an M6 star in figure \ref{JHvsHKOB1}, the region expected 
for objects with excess emission produced by hot dust in the innermost part of the disk
($\sim 0.1$ AU; \cite{mey97,muzerolle03}). 
The resulting inner-disk fraction for the entire sample (VLMSs AND BDs) 
is 0.38, and using the exact test for the success rate in binomial experiments 
(R-Statistical Software, \cite{ihaka_gentleman96}) we find a 95\% confidence 
level interval 0.17-0.64. These numbers are very similar to what we found from disk
accretion indicators (\ref{TTSfraction}), and consistent with the findings
reported by \cite{bri05}.

Five of the new members confirmed in this work classified as WTTS VLMSs 
with masses ($0.08<M/M_{\odot}<0.4$) are located in the 
$\sim 1\> deg^2$ OB1b field studied with the Spitzer Space Telescope by \cite{her07b}. 
Unfortunately, there are no similar observations for the objects 
we identified here in OB1a.
For the five OB1b objects we have photometry 
in the four IRAC bands with the exception of
05312373-0150245, not detected in the $[8.0]$ $\mu m$ band. These five members 
have all been classified as WTTS and Table 5 provides the IRAC photometry. 
Figure \ref{12vs34} shows the [3.6]-[4.5] vs. [4.5]-[5.8] and  
[3.6]-[4.5] vs. [5.8]-[8.0] color-color diagrams with the excess region
defined by \cite{luh05} (left panel) and the photospheric and CTTS locus
defined by \cite{har05} (right panel). All the objects exhibit photospheric colors
in both diagrams, consistent with their optical classification as non-accreting
objects; the exception is 05300324-0138428, classified as WTTS 
but whose [3.6]-[4.5] vs. [5.8]-[8.0] colors fall within the CTTS locus
defined by \cite{har05}, albeit offset toward smaller excess values.

Having only four objects in OB1b (all WTTS) with full IRAC photometry
prevents us from deriving a meaningful disk fraction; 
however, it is worth mentioning that for this same dataset
\cite{her07b} derived a disk frequency of 13\%,
using $\sim 100$ low-mass ($\sim 0.3\> M_\odot$) PMS stars.
Their determination agrees well with the CTTS fraction
derived by \cite{bri07b} in the same area. Our inner-disk 
fraction of $38^{+26}_{-21}$\% (see above), obtained
in an area $\sim 3$ times larger that of  the \cite{her07b},
is higher than their estimate, but consistent with the 
results from \cite{bri05}. As we have discussed in section \ref{TTSfraction},
the discrepancy could well be due, at least in part,
to the largely non-uniform
spatial distribution of disk-bearing CTTS in OB1.

In the young $\sigma$ Orionis cluster (age $\sim 3$ Myr) 
\cite{caballero07} combined optical, near-IR and Spitzer IRAC/MIPS measurements to
derive a disk fraction of $\sim 50$\% in a sample of 30 BDs down to 
$M\sim 0.015\> M_\odot$. However, this sample is a mixture 
of spectroscopically confirmed
members and candidates selected on the basis of photometry only. 
If we restrict the analysis to the 13 objects in their Table B.1 
which have spectral types, in the range M6-M7.5 (a spectral type 
range that overlaps the sample in this work), 
we find that only 5 objects exhibit excess IR emission
that can be attributed to circum(sub)stellar disks. 
This produces a somewhat lower disk fraction of 38\%, which agrees
very well with the value we found above, from the near-IR disk indicators. 

As in the case of the accretion-related indicators, our results from
the JHK 2MASS photometry suggest that: 1) Disk fractions are similar in VLMSs/BDs
and the lower mass PMS stars ($M\ga 0.3 M_\odot$) at a given age. 
2) The disk fraction 
near the substellar boundary falls off significantly, as
occurs for the lower mass stars ($0.8 \la M/M_\odot \la 0.3 $), 
during the period $\sim 4-8$ Myr. However, the present 
small size of the available samples precludes any detailed analysis on
how these fractions may evolve on both sides of the substellar limit.

We want to remark that strictly following the classification scheme 
by \citet{whi03}, of the 7 objects which show IR excesses, namely 
05300324-0138428, 05312373-0150245, 05350579-0121443, 05322069-0125511, 
05340726-0149380, 05345443-0144473 and 05390054-0150469, 
only the last two have been classified 
as CTTS.  Object 05300324-0138428 shows an $H\alpha$ emission equivalent width
at the WTTS/CTTS limit in Figure 7, and its IRAC colors place it at the
borderline between photospheric colors and IR excesses on both
color-color diagrams of Figure 9. A similar situation is that of
05312373-0150245, which is located near the WTTS/CTTS boundary in Fig.7,
given our 0.5 subclass uncertainty in the spectral type classification,
and its IRAC colors suggest it may have a small IR excess.
Objects 05322069-0125511 and 05340726-0149380 have $H\alpha$ emission 
equivalent widths well below the WTTS/CTTS boundary. We have IRAC photometry only for 
05322069-0125511, and its colors are clearly photospheric. 
In the near-IR JHK diagram (Fig. 8) both objects have the smallest $H-K$
values among the ``excess emission'' objects, lying closest
to the $H-K=0.45$ limit of our JHK excess box. 
BD 05350579-0121443 whose J, H and K magnitudes
are among the faintest of the sample, has an ``ABB'' 2MASS photometric 
flag which indicates a lower SNR in the H and K bands, and hence a larger
uncertainty in the $H-K$ color.  We conclude that, of the 5 objects
classified as WTTS but showing IR excess, 4 have $H\alpha$ equivalent widths,
from our low-resolution spectra,
that place them on or close to the border between WTTS and CTTS types. 
High resolution spectroscopy will be needed to analyze the $H\alpha$ line profile
and determine the presence of line wings extending to large velocities,
characteristic of accreting stars \citep[e.g.][]{muzerolle05}.

Of the 7 objects classified as CTTS only 2 fall within the IR excess box 
in Fig. 8. However, 3 out of the 5 CTTS outside the box have large enough 
$1-\sigma$ errorbars that would bring them into the excess emission region.
Therefore, within errors, 
only two CTTS really seem to lack IR excess emission at JHK. This is consistent
with findings for very low-mass stars in other star-forming regions like Taurus
\citep{luh03a}. Observations at longer wavelengths should allow
to detect excess IR emission in these two objects.

\section{Reddening and H-R diagrams}\label{extinctions_hrdiag}

The extinction towards each object was estimated from the $I-J$ color,
for which possible contributions from excess emission at short and long wavelengths
are minimized \citep{luh03b}. In
order to compute the intrinsic colors as a function of spectral type, we interpolated
the corresponding temperature from the spectral type-temperature relationship of 
\cite{luh03b}, into the \cite{bar98} models corresponding to ages of 7.9 Myr for
OB1a, and 3.2 Myr for OB1b. 

The resulting intrinsic colors and computed extinctions are
listed in Table \ref{table:data}. Extinctions are in the range 
$0.6<A_V<0.9$ for OB1a and $0.1<A_V<3.9$ for OB1b consistent
with previously reported values in both regions by \cite{bri05}.
Bolometric magnitudes were calculated using 
absolute $I$ band magnitudes $M_I$ and the bolometric correction 
$BC_I=0.02+0.575(V-I)-0.155(V-I)^2$ from \cite{le96}
where $V-I$ is the intrinsic color from \cite{kh95}. 
Figure 10 shows the resulting H-R diagrams and table 4 the mass 
estimates that result from the interpolation of luminosities 
and temperatures within the \cite{bar98} models. The mean ages 
derived are $\sim 4\pm 2Myr$ and $\sim 8\pm2 Myr$ for OB1a and
OB1b respectively, in good agreement with ages derived
based on low-mass stars \citep{bri05}.

\section{Summary and Conclusions}\label{summary_conclusions}

We have presented initial results of our search for 
substellar objects and VLMSs in the widely dispersed populations 
of the Orion OB1a and OB1b subassociations. Our results can 
be summarized as follows:

\begin{itemize}

\item Using deep optical I-band photometry and data from the 2MASS survey, we 
constructed optical-infrared color-color and color-magnitudes diagrams,
in which we selected candidates across $\sim$ $14.8$ $deg^2$ in Ori OB1 and over 
$\sim$  $6.7$ $deg^2$ in Ori OB1b with completeness down to $0.05M_{\odot}$ 
with $A_V\le 0.6$ for OB1a, and $0.072M_{\odot}$ with $A_V\le 2$ for OB1b. 
We obtained low resolution optical spectra for a subsample 
of 4 candidates over $\sim 0.8$ $deg^2$ of OB1a, and 26 candidates within
$\sim  2.5$ $deg^2$ of OB1b.

\item Through spectral signatures we confirmed 3 new 
members in OB1a, one of which is clearly substellar (spectral type M7, 
with an inferred mass $\sim 0.06M_{\odot}$) and 2 are VLMSs, one of which
is at the substellar limit (spectral type M6, with mass 
$M\sim0.072 M_{\odot}$). In OB1b we found 19 new members: 14 are VLMSs, 
of which 7 are at the substellar limit, and 5 are
substellar members with spectral types between M6.5 and M7.5.
From the spectroscopy we find that our photometric candidate 
selection technique is highly efficient (73\%) in picking out young, 
very low-mass and substellar members of the Orion OB1 association. 
Masses were estimated according to the \citep{bar98} models, and the less 
massive member (M7.5 spectral type), has an estimated mass 
$M\sim0.04M_{\odot}$.

\item The new members have been classified as CTTSs or WTTSs following the 
scheme of \cite{whi03}. We found that all three members confirmed in OB1a are 
WTTSs, while $39^{+25}_{-22}$\% of the members in OB1b are CTTSs.
These results are largely consistent with recent findings,
in the sense that the OB1a population is dominated by non-accreting
PMS stars, whereas the accretor fraction in the younger OB1b 
region can be estimated in the range $\sim 13 - 40$\% depending
on the exact location and area being considered, and the sample size.
Our findings indicate that in OB1b, 
the number of accretors in VLMSs/BDs is
similar to that derived for low-mass PMS 
stars (\cite{bri05}); also, that the overall number of accretors,
both in low-mass stars and among VLMSs/BDs falls off by a significant 
amount by ages $\sim 8-10$ Myr.

\item Of the 19 newly confirmed members of Ori OB1b, 7
(1 BD WTTS, 4 VLMSs WTTSs and 2 VLMSs CTTSs) exhibit excess emission 
in the 2MASS H-K color as would be expected of thermal emission
from hot dust in the innermost part of a circumstellar disk. 
None of the 3 Ori OB1a members show near-IR excesses.
We derive an inner-disk fraction of $38^{+26}_{-21}$\%, which is
in excellent agreement with our result for the accretor fraction.
Our inner-disk estimate for VLMSs and BDs
is also consistent with the results from \cite{bri05} for higher mass
stellar objects ($M\ga 0.3  M_{\odot}$), and with findings for BDs
spectroscopically confirmed in $\sigma$ Ori by \cite{caballero07}.
The few Ori OB1b members with Spitzer data are all WTTS, by
optical indicators; the IRAC photometry is largely consistent with
this classification. Only one object shows marginal indication of excess
emission at $8 \mu$m.

As in the case of the accretion-related indicators, the near-IR
indicators suggest that: 1) At a given age, the inner-disk fraction of objects at/near 
the substellar boundary is similar to that determined in the 
lower-mass ($M\ga 0.3  M_{\odot}$) PMS stars. 2) The inner-disk fraction
seems to fall off significantly, both in VLMSs/BDs and low-mass stars, 
once the population has aged to $\sim 8$ Myr,
as in Ori OB1a. However, the present 
small size of the available samples precludes any detailed analysis on
how these fractions may evolve on both sides of the substellar limit.

\item Finally we remark the discovery of two new members 05335219-0156398 (M7) and 
05390532-0135327 (M6) classified as CTTSs, without contamination by emission
lines from the cloud, both exhibiting strong 
H$\alpha$ emission, with $\rm W(H\alpha) \la -140${ \AA }, suggesting 
significant ongoing accretion.

\end{itemize}


\acknowledgments

This work has been supported in part by grant S1-200101144 of FONACIT, Venezuela. 
J. J. Downes, acknowledges support from grant 200400829 from FONACIT, Venezuela.
We thank Kevin Luhman for useful comments during the spectral classification, 
Nelson Caldwell for supplying us with the response 
curve of Hectospec, and Susan Tokarz, who is in charge of the reduction and 
processing of Hectospec spectra. 
This publication makes use of data products from the Two Micron All Sky
Survey, which is a joint project of the University of Massachusetts
and the Infrared Processing and Analysis Center/California
Institute of Technology, funded by the National Aeronautics and
Space Administration and the NSF. This research also has benefitted 
from the M, L, and T dwarf compendium housed at DwarfArchives.org and 
maintained by Chris Gelino, Davy Kirkpatrick, and Adam Burgasser,
from the ADS article retrieving service, and the Sistema de Colecci\'on de
Datos Observacionales SCDOBS (Ponsot et al. 2007).
We thank the assistance of the personnel, observers, telescope operators and 
technical staff at CIDA and FWLO, who made possible the observations at the 
J\"urgen Stock Schmidt-type telescope of the Venezuela National 
Astronomical Observatory (OAN) and at the MMT telescope at Fred Lawrence Whipple
Observatory (FLWO) of the Smithsonian Institution. Finally, we acknowledge
the comments from an anonymous referee, which helped improve
this article.

\clearpage

\clearpage
\begin{deluxetable}{cccccc}
\tablecolumns{6}
\tablewidth{0pt}
\tablecaption{Observation log of the individual scans used to produce the coadded scan}
\tablehead{\colhead{Date} & \colhead{Scan} & \colhead{$\alpha_i(J2000)$} & \colhead{$\alpha_f(J2000)$} &\colhead{Filters} & \colhead{Mean Seeing ('')}  }
\startdata
1998-12-13 & 504 & 04:10:00 & 05:51:00 & RBIV & 2.81 \\
1998-12-13 & 505 & 04:10:00 & 05:51:00 & RBIV & 2.95 \\
1998-12-29 & 501 & 04:10:00 & 05:51:00 & RBIV & 2.91 \\
1999-01-09 & 527 & 03:55:00 & 05:53:00 & RBIV & 2.93 \\
1999-01-09 & 528 & 03:55:00 & 05:53:00 & RBIV & 2.92 \\
1999-01-10 & 529 & 03:55:00 & 05:53:00 & RBIV & 2.85 \\
1999-01-09 & 528 & 03:55:00 & 05:53:00 & RBIV & 2.92 \\
1999-01-10 & 529 & 03:55:00 & 05:53:00 & RBIV & 2.85 \\
1999-01-10 & 530 & 03:55:00 & 05:53:00 & RBIV & 2.95 \\
1999-01-22 & 501 & 03:55:00 & 05:53:00 & RBIV & 3.06 \\
\enddata
\label{table:questlog}
\end{deluxetable}

\clearpage

\clearpage

\clearpage
\begin{deluxetable}{cccccc}
\tablecolumns{6}
\tablewidth{0pt}
\tablecaption{Hectospec observation log.}

\tablehead{\colhead{Date} & Subregion & \colhead{Field ID} & \colhead{$\alpha(J2000)$} & \colhead{$\delta(J2000)$} & \colhead{Integration [seg]} }
\startdata
Apr/09/2005  & OB1a& decm160\_04 & 05:17:59.67 & -01:35:19.4 & 2700\\ 
Nov/03/2004  & OB1b& decm160\_07 & 05:30:00.47 & -01:35:33.7 & 2700\\
Mar/04/2005  & OB1b& decm160\_08 & 05:33:55.64 & -01:34:29.8 & 2700\\ 
Nov/19/2004  & OB1b& decm160\_09 & 05:37:55.32 & -01:34:33.0 & 2700\\
\enddata
\label{table:HECTOSPEClog}
\end{deluxetable}

\clearpage
\begin{deluxetable}{ccc}
\tablecolumns{4}
\tablewidth{0pt}
\tablecaption{Spectral features from \cite{kir99} used for the semi-automated spectral classification.} 
\tablehead{
\colhead{Feature ID} & 
\colhead{$\lambda_c$ (\AA)} & 
\colhead{$\Delta \lambda $ (\AA)}
}
\startdata
TiO-1  & 4775 &30  \\
TiO-2  & 4975 &50  \\
TiO-3  & 5225 &75  \\
TiO-4  & 5475 &75  \\
TiO-5  & 5600 &50  \\
TiO-6  & 5950 &75  \\
TiO-7  & 6255 &50  \\
TiO-8  & 6800 &50  \\
TiO-9  & 7100 &50  \\
TiO-10  & 7150 &50 \\
VO-1   & 7460 &100 \\
VO-2   & 7840 &100 \\
VO-3   & 7940 &100 \\
\enddata
\label{table:molecfeatures}
\end{deluxetable}

\clearpage

\clearpage

\clearpage
\begin{deluxetable}{clllllllllclllllllll}
 \rotate
\tabletypesize{\tiny}
\tablecolumns{1}
\tablewidth{0pt}
\tablecaption{Catalog of very low-mass stars and brown dwarfs confirmed as new members of Orion OB1.}
\tablehead{ 
\colhead{2MASS-ID} &
\colhead{$R_C\pm\sigma R_C$} &
\colhead{$I_C\pm\sigma I_C$} &  
\colhead{$J$} &
\colhead{$H$} & 
\colhead{$K$} & 
\colhead{$ST\sigma ST$} & 
\colhead{adST} &
\colhead{W[H$\alpha$]} & 
\colhead{Lines} &
\colhead{WTTS/CTTS} &
\colhead{Av} &
\colhead{$Log(L/L_{\odot})$} & 
\colhead{$Teff$} & 
\colhead{$M(M_\odot)$} 
}
\startdata

05164635-0115233&19.225$\pm$0.047&17.121$\pm$0.010&14.946&14.415&14.127&M6.0$\pm$0.5&M6  & -26.8     &......&W&0.6&-1.81&2990& 0.07\\ 
05185578-0153232&20.204$\pm$0.100&17.767$\pm$0.015&15.364&14.640&14.260&M7.0$\pm$1.0&M7  & -18.5     &......&W&0.5&-1.80&2880& 0.06\\ 
05193522-0144501&18.706$\pm$0.023&16.882$\pm$0.007&14.864&14.330&14.020&M4.7$\pm$0.5&M4.5& -13.3     &......&W&0.9&-1.79&3197& 0.15\\ 
05300324-0138428&....................&16.272$\pm$0.013&13.419&12.515&11.895&M2.7$\pm$0.5&M3&-20.0    &......&W&3.9&-0.64&3415& 0.40\\
05311727-0136587&19.049$\pm$0.037&16.989$\pm$0.008&14.944&14.432&14.078&M6.0$\pm$0.5&M6  & -16.8     &......&W&0.1&-1.63&2990& 0.08\\ 
05312373-0150245&19.967$\pm$0.066&17.839$\pm$0.016&15.643&15.104&14.485&M6.2$\pm$0.5&M6  & -24.4     &......&W&0.6&-1.85&2990& 0.07\\ 
05322069-0125511&18.376$\pm$0.015&16.569$\pm$0.005&14.656&14.205&13.709&M4.8$\pm$0.5&M5  &  -8.8     &......&W&0.2&-1.56&3125& 0.13\\ 
05323839-0127204&....................&18.059$\pm$0.018&15.800&15.298&14.986&M5.5$\pm$0.5&M5.5&-26.0  &......&C&1.0&-1.91&3057& 0.09\\ 
05332365-0136234&20.081$\pm$0.071&17.819$\pm$0.014&15.379&14.856&14.458&M6.0$\pm$1.0&M6  & -22.3     &......&W&1.3&-1.67&2990& 0.08\\ 
05333251-0112519&18.158$\pm$0.024&16.240$\pm$0.009&14.013&13.347&12.921&M5.0$\pm$1.0&M5  & -31.1     &......&C&1.2&-1.19&3125& 0.17\\ 
05335219-0156398&....................&18.102$\pm$0.024&15.736&14.967&14.579&M7.0$\pm$1.5&M7 &-207.4  &......&C&0.3&-1.73&2880& 0.06\\ 
05340726-0149380&20.148$\pm$0.100&17.876$\pm$0.017&15.641&15.271&14.766&M5.5$\pm$1.0&M5.5& -10.4     &......&W&1.0&-1.83&3057& 0.09\\ 
05345443-0144473&17.020$\pm$0.005&15.773$\pm$0.004&13.381&12.511&11.901&M3.0$\pm$0.5&M3  & -32.8     & HeI  &C&2.5&-0.78&3415& 0.40\\ 
05350579-0121443&....................&18.928$\pm$0.040&16.341&15.835&15.229&M6.5$\pm$1.0&M6.5&-18.1  &......&W&1.4&-2.00&2935& 0.07\\ 
05370790-0129204&20.344$\pm$0.090&18.173$\pm$0.017&15.869&15.564&15.216&M4.8$\pm$0.5&M5       &  -4.7&......&W&1.4&-1.91&3125& 0.13\\ 
05372198-0129140&18.676$\pm$0.020&16.484$\pm$0.005&14.209&13.703&13.333&M6.1$\pm$0.5&M6       & -26.8&......&W&0.8&-1.26&2990& 0.10\\ 
05373853-0124270&18.617$\pm$0.026&16.701$\pm$0.007&14.592&14.037&13.625&M6.2$\pm$0.5&M6       & -32.4&......&W&0.3&-1.47&2990& 0.09\\ 
05374145-0155519&....................&18.295$\pm$0.031&15.987&15.484&15.115&M7.0$\pm$0.5&M7   & -47.0& HeI  &C&0.2&-1.83&2880& 0.06\\ 
05380232-0147405&18.406$\pm$0.020&16.249$\pm$0.004&14.088&13.487&13.103&M6.5$\pm$0.8&M6.5     & -16.5&......&W&0.1&-1.24&2935& 0.08\\ 
05390054-0150469&19.613$\pm$0.065&17.732$\pm$0.018&15.476&15.020&14.539&M6.2$\pm$1.0&M6       & -67.1&......&C&0.7&-1.78&2990& 0.07\\ 
05390532-0135327&19.604$\pm$0.041&17.619$\pm$0.011&15.356&14.784&14.454&M5.8$\pm$0.5&M6&-140.3       & HeI  &C&0.8&-1.72&2990& 0.07\\ 
05390910-0200267&....................&19.102$\pm$0.086&16.166&15.559&15.413&M7.7$\pm$1.0&M7.5& -93.5 &......&?&1.6&-1.90&2795& 0.04\\ 

\enddata

\label{table:data}
\end{deluxetable}

\clearpage
\begin{figure}
\plotone{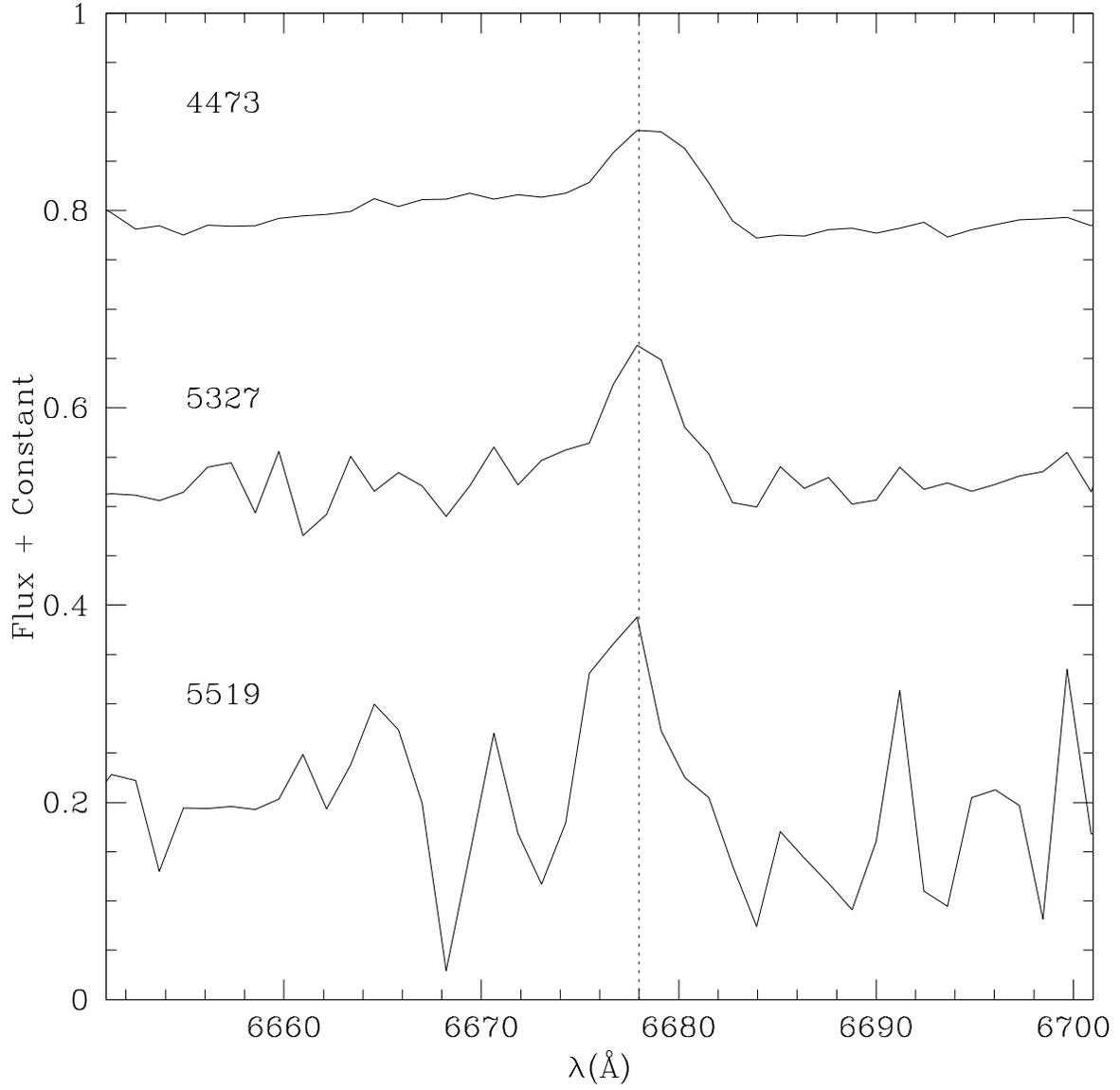}
\caption{HeI emission lines observed in 05345443-0144473, 05390532-0135327 and 05374145-0155519 
supporting their classification as CTTS.}
\end{figure}

\clearpage
\begin{figure}
\plotone{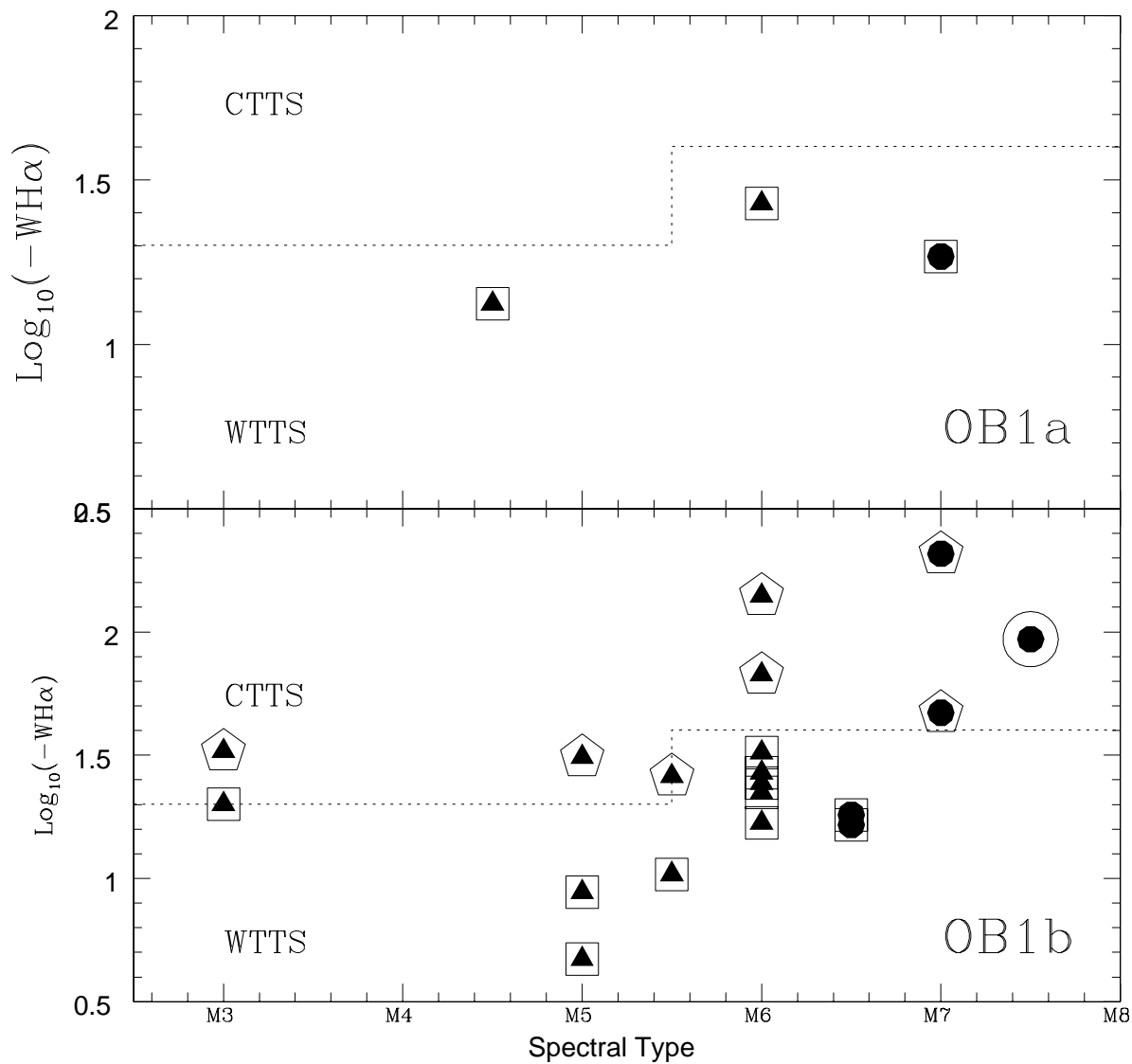}
\caption{$\rm W(H\alpha)$ vs. spectral type for the new members. 
Solid circles and triangles indicate BDs and VLMSs respectively, 
confirmed as new members. Squares represent members classified as WTTSs and 
pentagons indicate members classified as CTTSs. The dotted line indicates 
the separation between CTTSs and WTTSs from \cite{whi03}. The circled dot 
indicates member 05390910-0200267, that could not be classified as either
WTTS or CTTS as explained in subsection 7.1.} \label{wha_ij}.
\end{figure}

\clearpage
\begin{figure}
\plotone{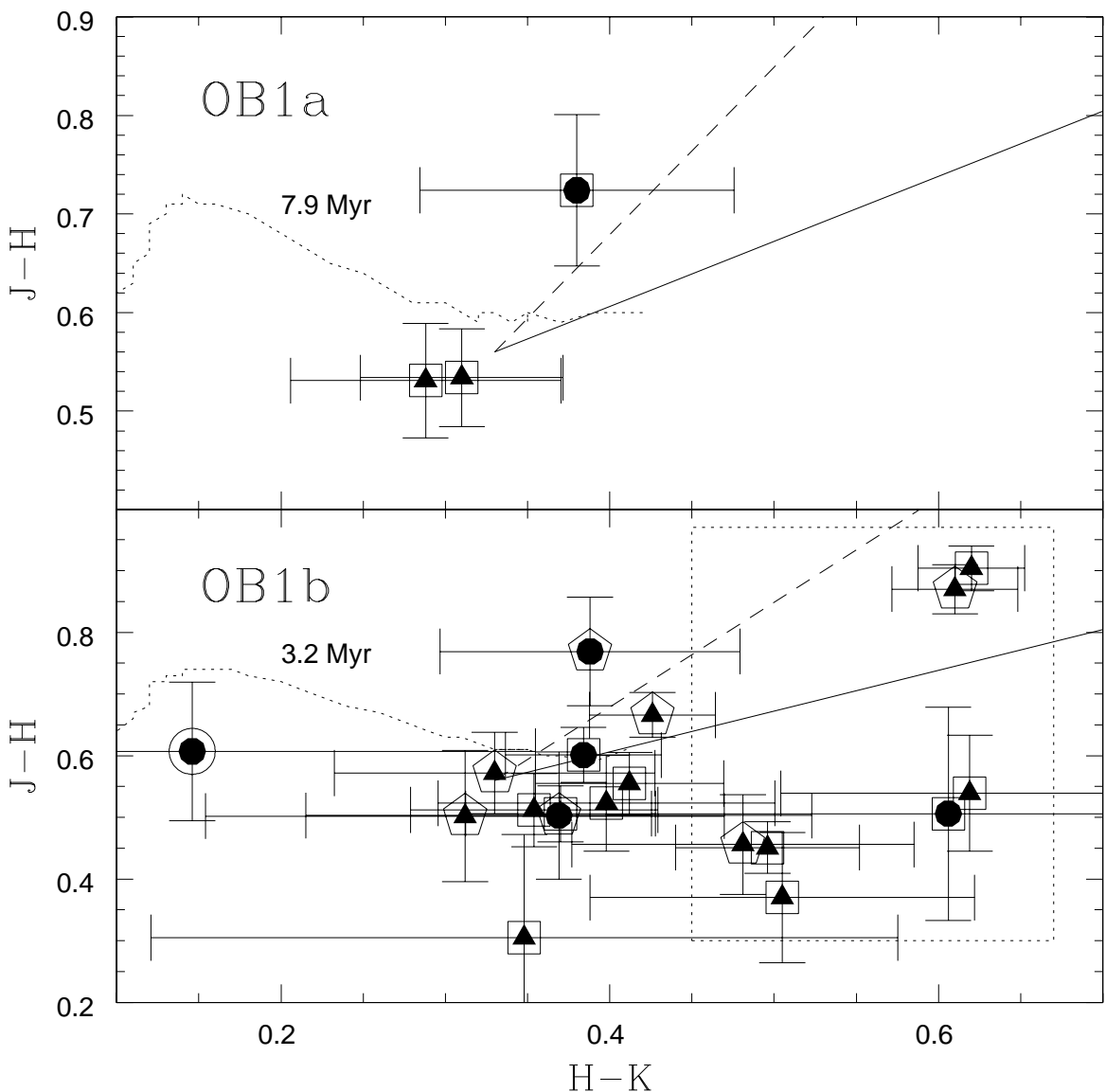}
\caption{Observational $J-H$ vs. $H-K$ diagram of the OB1a and OB1b members
confirmed at this work. Symbols are as in Figure \ref{wha_ij}. The dotted lines
represent the 7.9 Myr and 3.2 Myr isochrones from \cite{bar98}, that correspond
to the ages found by \cite{bri05} on the basis of low mass stars in each sub association.
The dashed line represents the reddening line for an M6 spectral type, and the dotted box
encloses objects showing H-K excess. The solid line is
an extrapolation of the CTTS locus from \cite{mey97} for early M stars
to an M6 spectral type.}\label{JHvsHKOB1}
\end{figure}

\clearpage
\begin{deluxetable}{cllclllllcll}
\tabletypesize{\tiny}
\tablecolumns{1}
\tablewidth{0pt}
\tablecaption{IRAC photometry of the new VLM members of OB1b}
\tablehead{
\colhead{ID-2MASS}&
\colhead{STad} &
\colhead{WHa} &
\colhead{WTTS/CTTS} &
\colhead{[3.6]} &
\colhead{$\sigma$[3.6]} &
\colhead{[4.5]} &
\colhead{[$\sigma$4.5]} &
\colhead{[5.8]} &
\colhead{[$\sigma$5.8]} &
\colhead{[8.0]} &
\colhead{[$\sigma$8.0]}
}
\startdata
05300324-0138428 & M3   & -20.0 &  W & 11.180 & 0.003 & 10.922& 0.003 & 10.785& 0.008 & 10.273 & 0.008 \\
05311727-0136587 & M6   & -16.8 &  W & 13.708 & 0.009 & 13.635& 0.012 & 13.684& 0.054 & 14.297 & 0.239 \\
05312373-0150245 & M6   & -24.4 &  W & 14.464 & 0.012 & 14.223& 0.017 & 13.995& 0.059 & ......... & ........ \\
05322069-0125511 & M5   &  -8.8 &  W & 13.528 & 0.008 & 13.458& 0.011 & 13.501& 0.048 & 13.523 & 0.090 \\
05323839-0127204 & M5.5 & -26.0 &  W & 14.611 & 0.014 & 14.570& 0.022 & 14.533& 0.100 & 15.197 & 0.420 \\
\enddata

Note: Photometry is from \cite{her07b}. Spectral types, W(H$\alpha$) and 
WTTS/CTTS classification are from our Table \ref{table:data}
\label{table:datairac}
\end{deluxetable}

\clearpage
\begin{figure}
\plotone{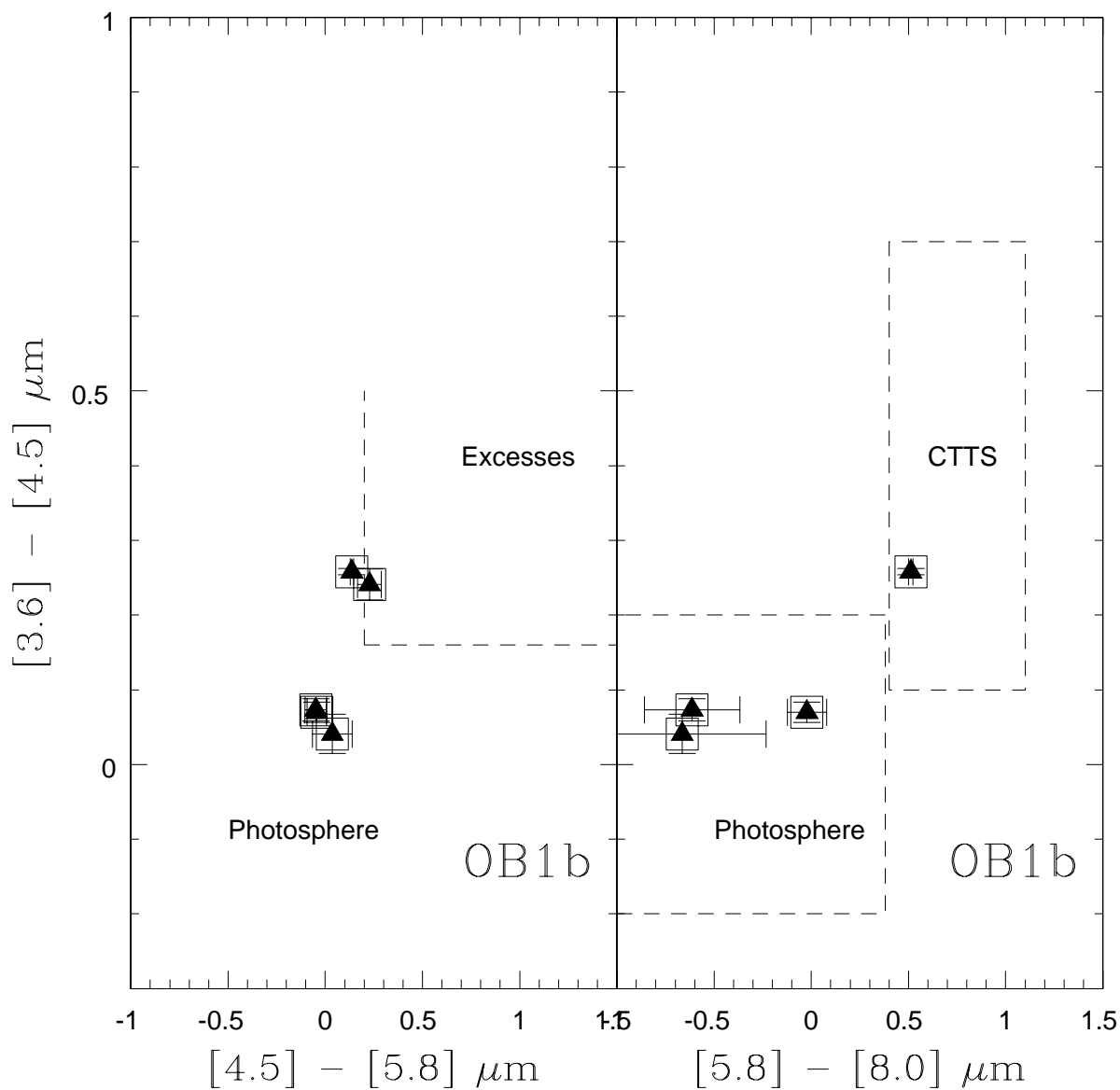}
\caption{Color-color diagrams of the five OB1b members with available IRAC photometry. The dashed lines 
in the left panel indicate the region of excess IR emission
defined by \cite{luh05}. Dashed boxes
in the right panel indicate the position of the photospheric locus and 
the Class II source locus from  \cite{har05}. Symbols are as in Figure \ref{JHvsHKOB1}.}\label{12vs34}
\end{figure}

\clearpage
\begin{figure}
\plotone{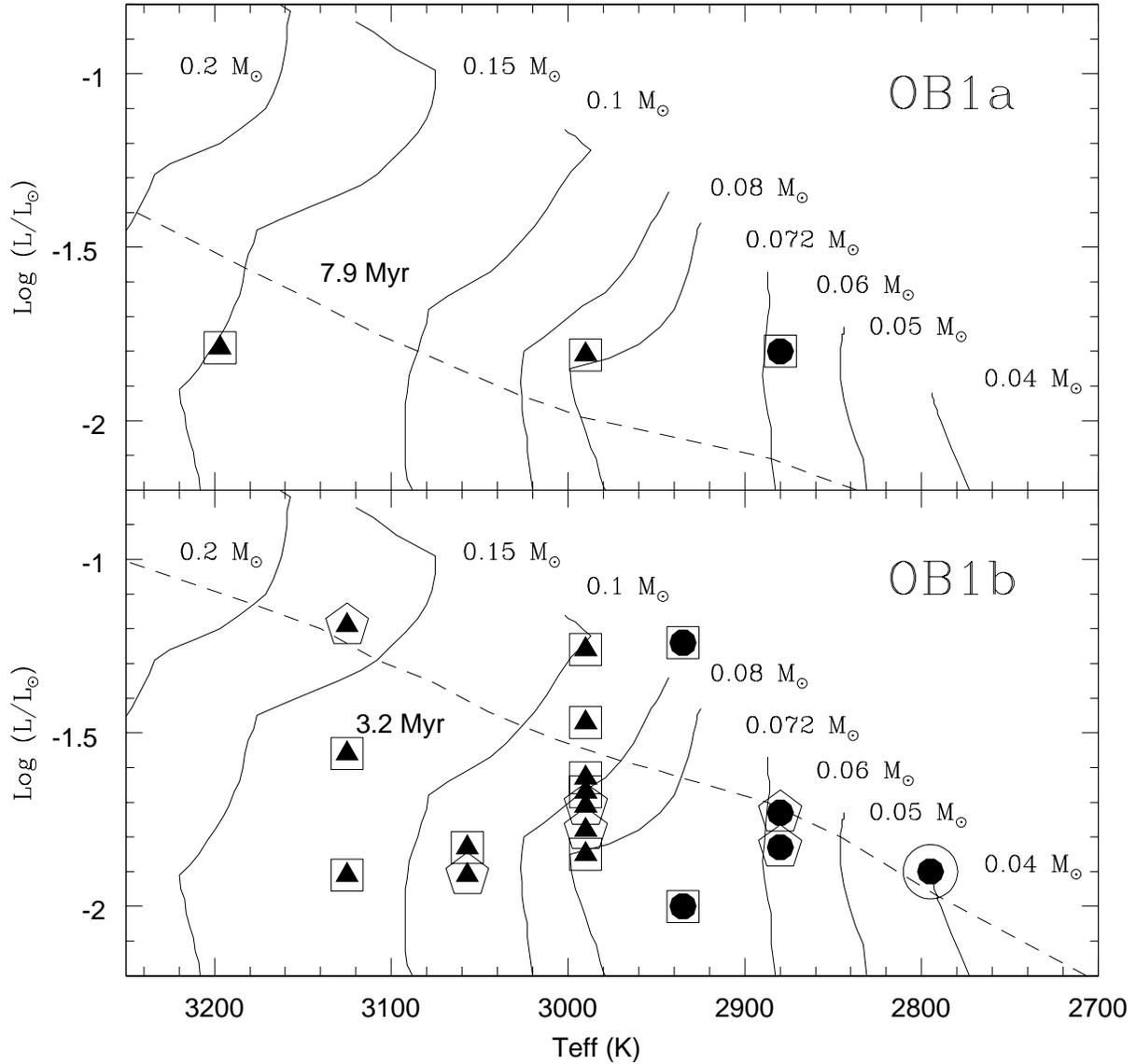}
\caption{H-R diagrams of confirmed members located 
in Hectospec OB1a and OB1b fields. Symbols are as in Figure \ref{wha_ij}. 
Evolutionary tracks and isochrones 
are from \cite{bar98} where the 3.2 Myr and 7.9 Myr isochrones correspond to 
the ages of OB1a and of OB1b respectively, computed from the low mass members 
($M>0.3M_\odot$) in color-magnitude 
diagrams by \citep{bri05}.}\label{derIvsIJOB1}
\end{figure}

\end{document}